\documentclass[]{elsart}

\usepackage{epsfig}
\usepackage{graphics}
\usepackage{graphicx}
\usepackage[centertags]{amsmath}
\usepackage{multicol}
\usepackage{captcont}
\usepackage{url}
\usepackage{subfigure}
\usepackage{algorithmic}

\begin{document}

\begin{frontmatter}

\title{BIO-DEVELOPMENT OF MOTORWAY NETWORKS IN THE NETHERLANDS: \\A SLIME MOULD APPROACH\\ \vspace{0.5cm}
Final version of this paper is published in\\
Advances in Complex Systems (2012) \\ DOI: 10.1142/S0219525912500348
}

\author{Andrew Adamatzky$^1$,Michael Lees$^2$ and Peter M.A. Sloot$^{2,3,4}$} 

\address{$^1$~University of the West of England, Bristol BS16 1QY, United Kingdom \\ 
{\tt andrew.adamatzky@uwe.ac.uk}\\
$^2$~ Nanyang Technological University, Singapore\\
{\tt mhlees@ntu.edu.sg} \\
$^3$~ University of Amsterdam, Amsterdam, The Netherlands\\
$^4$~ National Research University ITMO, Russia\\
{\tt p.m.a.sloot@uva.nl} 
}
\date{}

\maketitle

\begin{abstract}
\noindent
Plasmodium of acellular slime mould \emph{Physarum polycephalum} is a
very large eukaryotic microbe visible to the unaided eye. During its
foraging behaviour the plasmodium spans sources of nutrients with a
network of protoplasmic tubes. In this paper we attempt to address the
following question: is slime mould capable of computing transport
networks? By assuming the sources of nutrients are cities and
protoplasmic tubes connecting the sources are motorways, how well does
the plasmodium approximate existing motorway networks?  We take the
Netherlands as a case study for bio-development of motorways, while it
has the most dense motorway network in Europe, current demand is
rapidly approaching the upper limits of existing capacity. We
represent twenty major cities with oat flakes, place plasmodium in
Amsterdam and record how the plasmodium spreads between oat flakes via
the protoplasmic tubes. First we analyse slime-mould-built and
man-built transport networks in a framework of proximity graphs to
investigate if the slime mould is capable of computing existing
networks. We then go on to investigate if the slime mould is able
calculate or adapt the network through imitating restructuring of the
transport network as a response to potential localalized flooding of the
Netherlands.

\vspace{0.5cm}

\end{abstract}

\end{frontmatter}

\section{Introduction}

The approximation or computation of shortest path transportation
networks has drawn significant attention from the field of
Unconventional Computing Sciences. Nature-inspired computing paradigms
and experimental implementations have been successfully applied to
calculation of a minimal-distance path between two given points in a
space or a road network. Computational models of ant-based
optimisation have been shown to be an effective way of developing
novel approaches towards load-balancing of
telecommunications~\cite{Dorigo_2004}, which indeed involves dynamical
design of transport links for packets. Other work includes a
shortest-path problem solved in experimental reaction-diffusion
chemical systems~\cite{adamatzky_2005}, gas-discharge analog
systems~\cite{reyes_2002}, spatially extended crystallization
systems~\cite{adamatzky_hotice}, formation of fungi mycelian
networks~\cite{jarret_2006} and plasmodium of \emph{Physarum
  polycephalum}~\cite{nakagaki_2001a}.

Amongst all experimental prototypes of path-computing devices slime
mould \emph{Physarum polycephalum} is perhaps the most cost efficient
biological substrate available, coupled with fact that it is both easy to
cultivate and observe, it makes an excellent computational substrate.
These are the main reasons we adopt it for this work.

Acellular slime mould \emph{Physarum polycephalum} has quite a
sophisticated life cycle~\cite{stephenson_2000}, which includes
various stages such as: fruit bodied, spores, single-cell amoebas, and
syntsyncytium. Plasmodium is a vegetative stage of \emph{Physarum
  polycephalum}, it is a syncytium, a single cell, where many nuclei
share the same cytoplasm. The plasmodium consumes microscopic
particles, and during its foraging behaviour the plasmodium spans
scattered sources of nutrients with a network of protoplasmic
tubes. The protoplasmic network is usually optmized to cover all
sources of food while still managing to guarantee robust and quick
distribution of nutrients in the plasmodium body. Plasmodium's
foraging behaviour can be interpreted as computation, with data
represented by spatial distribution of attractants and repellents, and
results represented by the structure of protoplasmic
networks~\cite{adamatzky_physarummachines}.  Plasmodium is capable of
solving computational problems with natural parallelism, namely
shortest path~\cite{nakagaki_2001a} and hierarchies of planar
proximity graphs~\cite{adamatzky_ppl_2009}, computation of plane
tessellations~\cite{shirakawa}, execution of basic logical computing
schemes~\cite{tsuda_2004,adamatzky_gates}, and natural implementation
of spatial logic and process
algebra~\cite{schumann_adamatzky_2009}. For further examples see the
overview of Physarum-based computers in
\cite{adamatzky_physarummachines}.

In previous work~\cite{adamatzky_UC07} we have evaluated the
road-modeling potential of \emph{P. polycephalum}, however, previous
results were inconclusive.  A step forward biological-approximation,
or evaluation, of man-made road networks was done in our previous
work on approximation of United Kingdom motorways and Mexican
Federal highways by plasmodium of \emph{Physarum
  polycephalum}~\cite{adamatzky_jones_2009,adamatzky_mexican}. In
both cases it was shown that transportation links constructed by
plasmodium sufficiently determines man-made motorways, with some pernicious differences.
Comparing results for United Kingdom and Mexico we found that shape of
a country and spatial configuration of urban areas or cities
sufficiently determines behaviour of the plasmodium. More experiments
are necessary to provide generalisation, in order to develop a theory
of slime-mould based road planning and urban development.

In this paper we hope to move towards a more general understanding of
slime moulds capability to compute road networks by investigating the
roads in the Netherlands. The Netherlands presents an excellent case
study as it has the highest density motorway network in
Europe. Moreover, the demand on the system is at levels which are
reaching current limits, with a total length of $132,397km$ and usage
of $140 \times 10^9$ people per km per
yea\footnote{www.autosnelwegen.nl}.  Such high-occupancy may pose a
need for urgent expansion of the transport networks and a better
understanding of the limitations to that growth. The Netherlands is
also at risk of significant
flooding\footnote{http://urbanflood.eu/default.aspx}.

The remainder of this paper is structured as follows. We delinate the
experimental method and setup in
section~\ref{methods}. Section~\ref{results} presents the principal
experimental results, which are then analysed in a framework of
proximity graphs in section~\ref{proximitygraphs}. Restructuring of
Physarum-approximated transport links for the case of partial flooding
of the Netherlands is described in section~\ref{flooding}. The paper
then concludes with a summary of the work and ideas for further
studies in section~\ref{discussion}.

\section{Methods}
\label{methods}
All experiements are conducted with Plasmodium of \emph{
  P. polycephalum} that is cultivated in a plastic container. The
Plasmodium are first placed on paper kitchen towels, sprinkled with
still water and fed with oat flakes\footnote{Asda's Smart Price
  Porridge Oats}. The experiments are conducted in $120 \times 120$~mm
polyestyrene square Petri dishes with rounded corners. The Plasmodium
will eventually grow on Agar plates, which are cut into the shape
of the Netherlands. The Agar plates are formed using 2\% agar gel
(Select agar, Sigma Aldrich).

\begin{figure}[!tbp]
\centering
\subfigure[]{\includegraphics[width=0.5\textwidth]{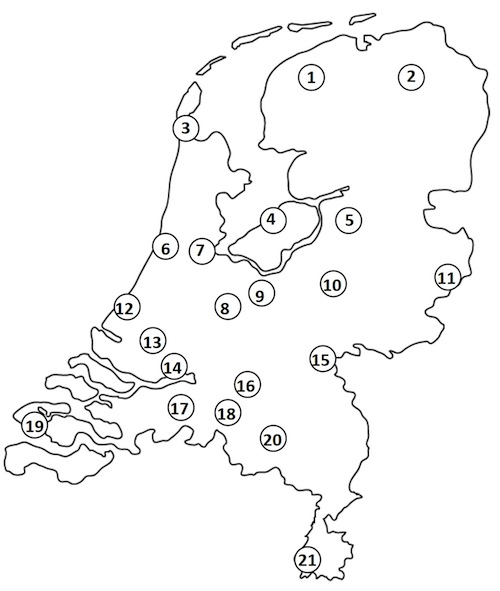}}
\subfigure[]{\includegraphics[width=0.49\textwidth]{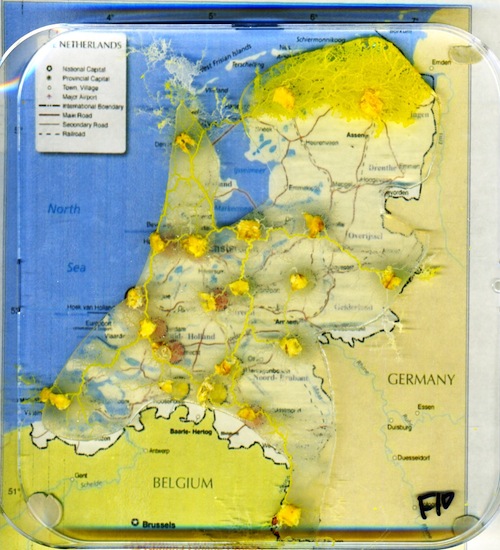}}
\subfigure[]{\includegraphics[width=0.49\textwidth]{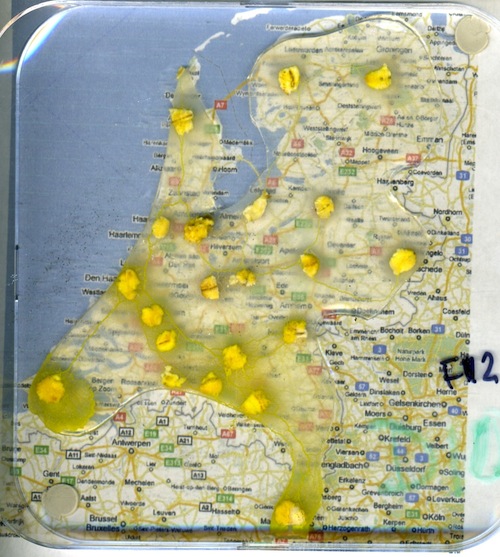}}
\subfigure[]{\includegraphics[width=0.49\textwidth]{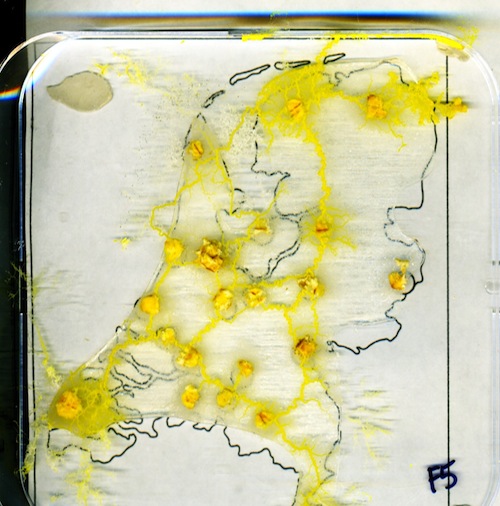}}
\caption{Experimental basics. 
(a)~Outline map of  the Netherlands with twenty one sources of nutrients indicated.
(b)--(d)~Snapshots typical setups: urban areas are represented by oat flakes, plasmodium 
is inoculated in Amsterdam, the plasmodium spans oat flakes by protoplasmic transport 
network.}
\label{urbanareas}
\end{figure}

All experiments consider the twenty one most populous urban areas in
the Netherlands (Fig.~\ref{urbanareas}a):
\begin{multicols}{2}
\begin{enumerate}
\item Leeuwarden % 1
\item Groningen %2 
\item Den Helder %3
\item Lelystad %4
\item Zwolle %5
\item Haarlem %6
\item Amsterdam %7
\item Utrecht %8
\item Amersfoort %9
\item Apeldoon %10 
\item Enschede %11
\item Den Haag %12
\item Rotterdam %13
\item Dordrecht %14
\item Nijmegen %15
\item Hertogenbosch %16
\item Breda %17
\item Tilburg %18 
\item Middelburg %19
\item Eindhoven %20 
\item Maastricht. %21
\end{enumerate}
\end{multicols}

\vspace{0.5cm}

Further we refer to the urban regions as $\mathbf U$. The regions in $\mathbf U$ are projected onto the
gel in the following manner: oat flakes are placed in the positions of each region (Fig.~\ref{urbanareas}b). 
At the beginning of each experiment a piece of plasmodium, usually already attached to an oat flake, is 
placed in Amsterdam (region 7 in Fig.~\ref{urbanareas}a).

The Petri dishes with plasmodium are kept in darkness, at a
temperature of between 22 and 25~C$^{\text o}$, except for short
periods of observation and image recording. Periodically the dishes
are scanned using an Epson Perfection 4490 scanner. Scanned images of
dishes are enhanced to increase readability of the image, this is done
by increasing saturation and contrast (saturation is increased to 55
and contrast to 40). A total of 62 experiments were conducted.

To ease understading of experimental images we provide complementary
binary version of each image, where appropriate. In these images each
pixel of the color image is assigned a black color if red $R$ and green
$G$ components of its RGB color exceed some specified thresholds, $R >
\theta_R$, $G > \theta_G$ and the blue component $B$ does not exceed some
threshold value $B < \theta_B$; otherwise, the pixel is assigned a white
color (exact values of the thresholds are indicated in the figure
captions as $\Theta=(\theta_R, \theta_G, \theta_B)$).

\section{Transport links via foraging}
\label{results}
In the following we present experimental results which show the the
plasmodium is capable of computing, or calculating, the transport
links between each of the twenty one most populus areas of the
Netherlands.

\begin{figure}[!tbp]
\centering
\subfigure[$t=$12~h]{\includegraphics[width=0.4\textwidth]{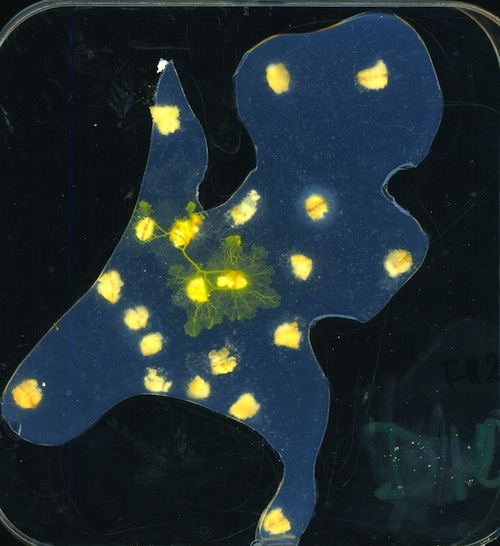}}
\subfigure[$t=$34~h]{\includegraphics[width=0.4\textwidth]{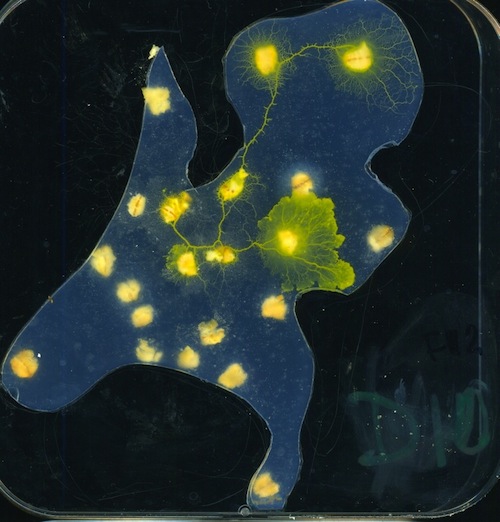}}
\subfigure[$t=$57~h]{\includegraphics[width=0.4\textwidth]{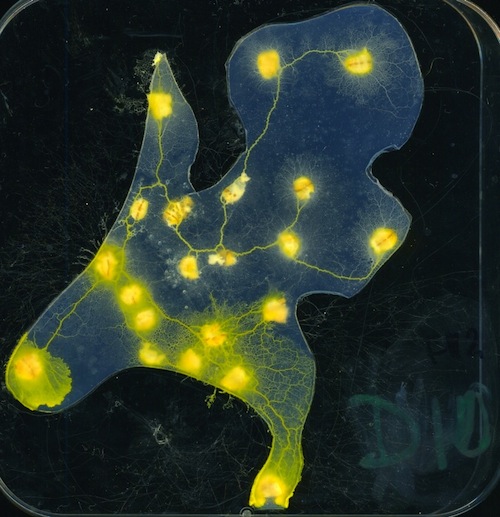}}
\subfigure[$t=$12~h]{\includegraphics[width=0.4\textwidth]{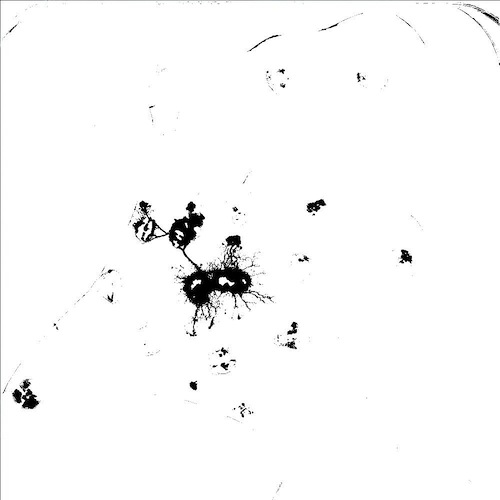}}
\subfigure[$t=$34~h]{\includegraphics[width=0.4\textwidth]{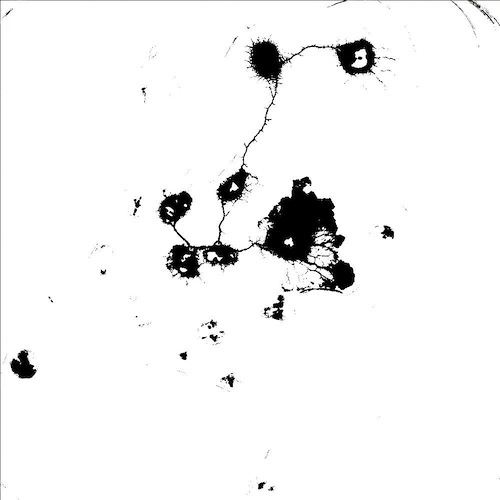}}
\subfigure[$t=$57~h]{\includegraphics[width=0.4\textwidth]{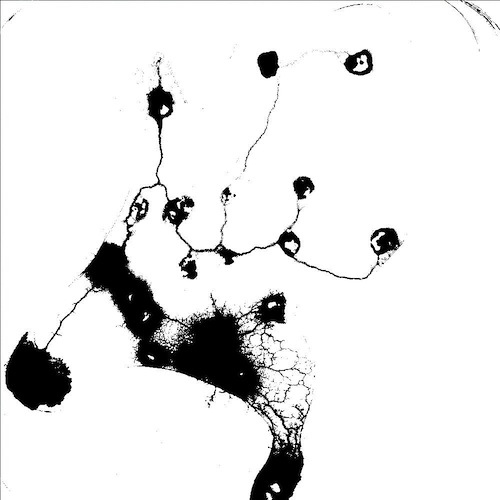}}
\caption{Illustrative example of plasmodium development on configuration of cities represented by oat flakes:
(a)--(c)~scanned image of experimental Petri dish. Time elapsed from inoculation is shown in the sub-figure captions.
(d)--(f)~binary images, $\Theta=(100,100,100)$. 
}
\label{f12}
\end{figure}

\begin{figure}[!tbp]
\centering
\subfigure[]{\includegraphics[width=0.44\textwidth]{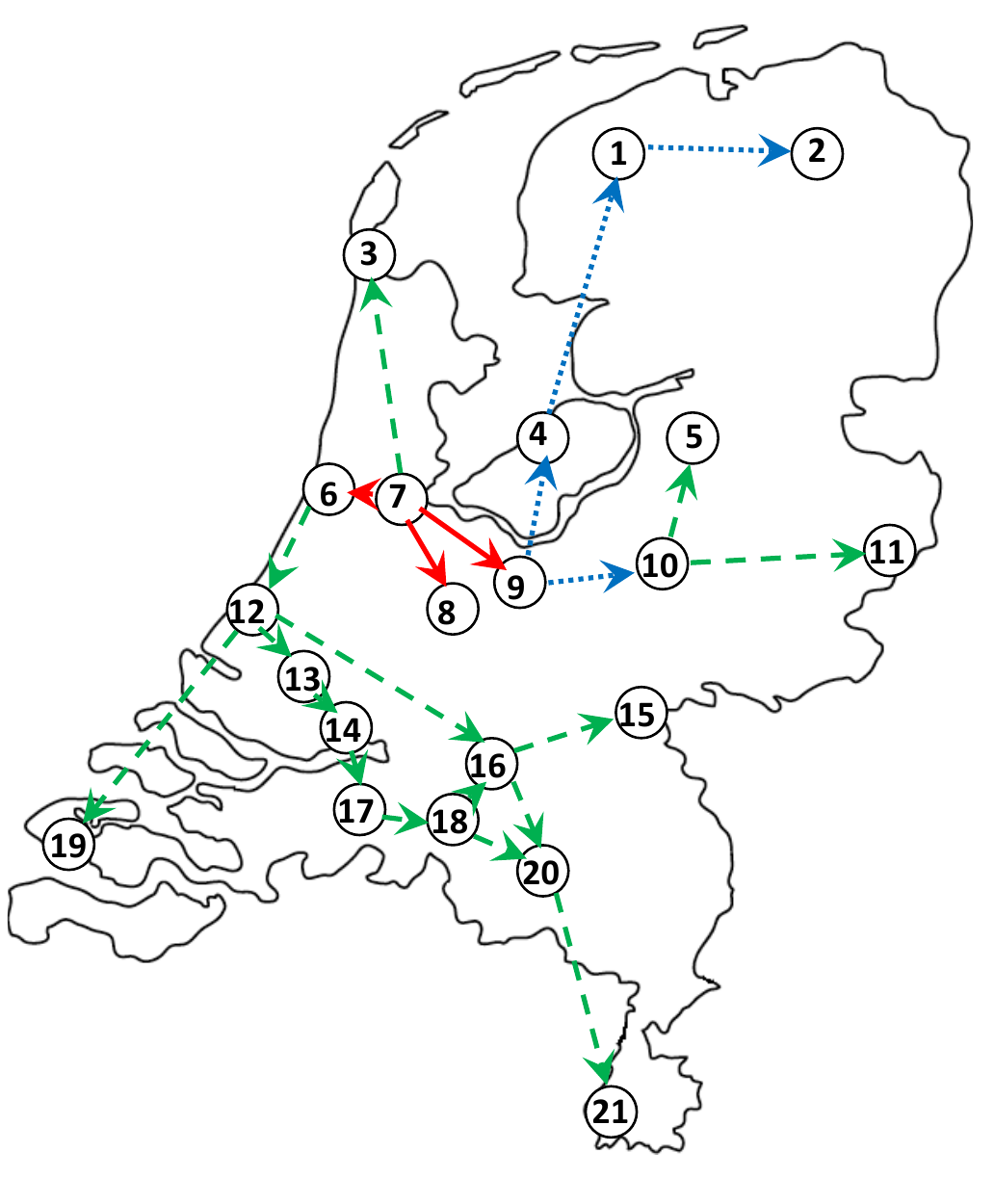}}
\subfigure[]{\includegraphics[width=0.53\textwidth]{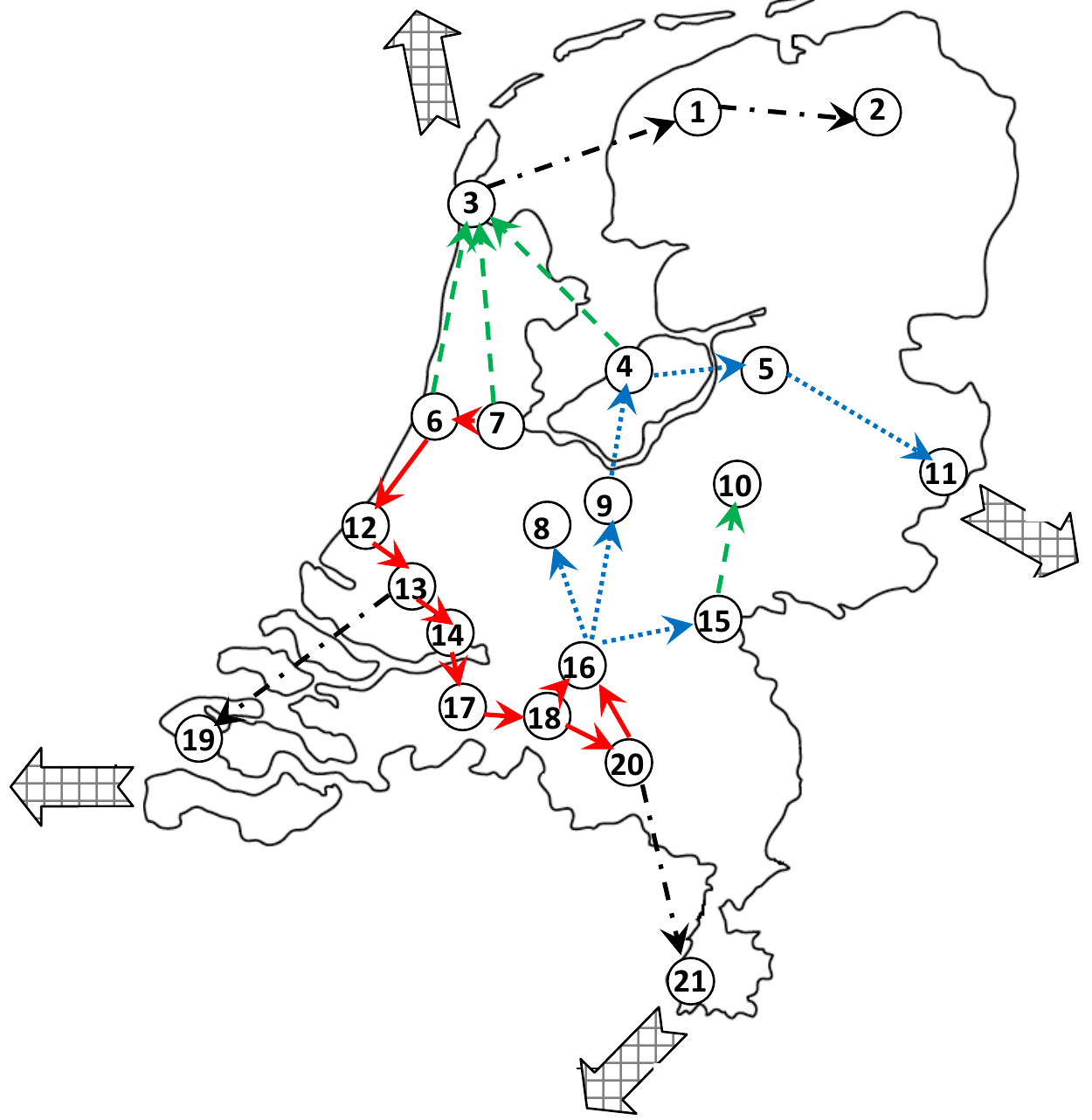}}
\caption{Diagram of colonisation dynamics derived from experiments Fig.~\ref{f12}~(a) and Fig.~\ref{f10}~(b): 
links developed in 12 hours after inoculation are shown by red solid arrows, 
in 34 hours by blue dotted arrows, in 57 hours by green dashed lines, in 80 hours by dash-dotted lines. Large mesh-patterned arrows
indicate migration of plasmodium outside the country.}
\label{arrowsf12}
\end{figure}

In a laboratory experiment, illustrated in Fig.~\ref{f12}, the
following chain of events unfolds (dynamics of colonisation is
schematically represented in Fig.~\ref{arrowsf12}).  An oat flake
colonised by plasmodium was placed on top of the oat flake
representing Amsterdam. In 12 hours the plasmodium follows gradients
of chemoattractants, links Amsterdam with Haarlem, and propagates
towards Utrecht and Amersfoort, spreading in all directions except
north-west (Fig.~\ref{f12}ad). After 34 hours the plasmodium colonizes
Leeuwarden and Groningen. It develops clearly visible protoplasmic
tubes, which represent a transport link Amersfoort - Lelystad -
Leeuwarden - Groningen (Fig.~\ref{f12}be). In the same time interval
the plasmodium colonises Apeldoon and start propagations towards
Zwolle and Enschede (Fig.~\ref{f12}be).

After a total of 57 hours the plasmodium connects Apeldoon with
Enschede and Zwolle by protoplasmic tubes and colonised south-west
part of the country. Namelym, the plasmodium links Haarlem and The
Hague and builds a route from The Hague to Middelburg and a link Hague
- Rotterdam - Dordrecht - Breda - Tilburg - Hertogenbosch
(Fig.~\ref{f12}cf, Fig.~\ref{arrowsf12}a, green dashed lines). At the
same time the plasmodium forms a protoplasmic tube directly connecting
Amsterdam Den Helder, and The Hague with Hertogenbosch, and develops the
links Hertogenbosch - Nijmegen and {Tilburg - Hertogenbosch} -
Eindhoven - Maastricht (Fig.~\ref{f12}cf).

We observe that the dynamics of colonisation is non-uniform
(Fig.~\ref{arrowsf12}a). The Plasmodium does not spread or diffuse in
all directions simultaneously but rather colonises north-north-west
part of the country first and only then explores south-south-west.
This may be due to the fact that centres of activity
(biochemical oscillators) form during propagation, and the contractive
waves evoked by the oscillators that force the protoplasm to move towards the
oscillators. Therefore, if an oscillator is formed in the north part of
plasmodium, then propagation in all other directions would be
suppressed.

\begin{figure}[!tbp]
\centering
\subfigure[$t=$12~h]{\includegraphics[width=0.4\textwidth]{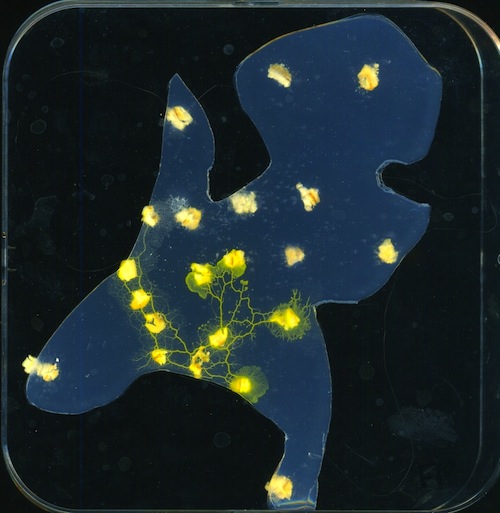}}
\subfigure[$t=$34~h]{\includegraphics[width=0.4\textwidth]{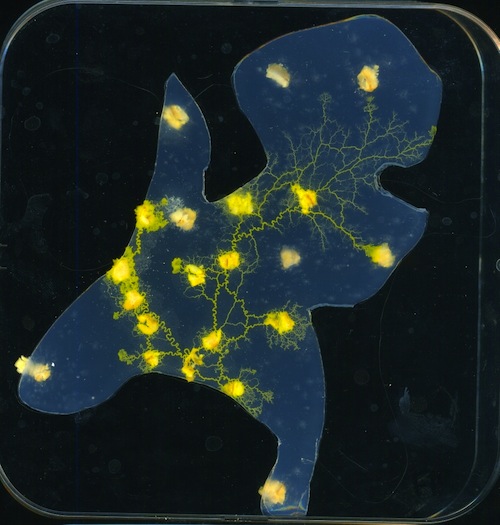}}
\subfigure[$t=$57~h]{\includegraphics[width=0.4\textwidth]{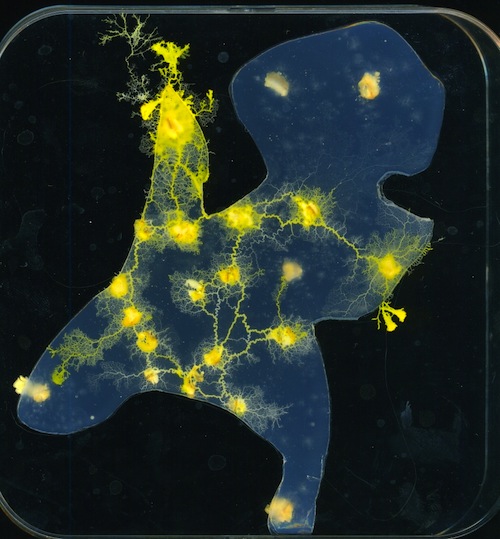}}
\subfigure[$t=$80~h]{\includegraphics[width=0.4\textwidth]{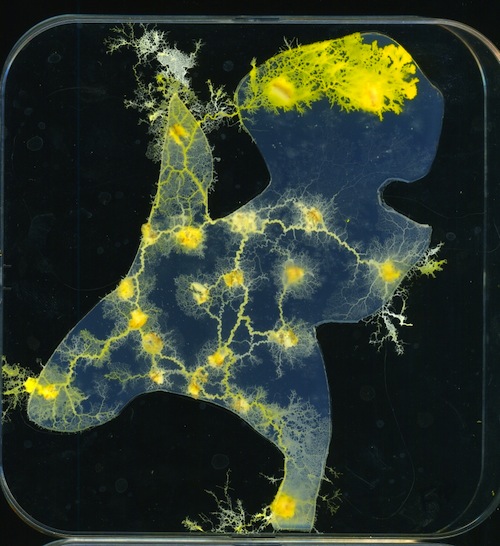}}
\captcont{Plasmodium spreads beyond `dedicated' experimental domain:
(a)--(d)~scanned image of experimental Petri dish. Time elapsed from inoculation is shown in the sub-figure captions.
(e)--(h)~binary images, $\Theta=(100,100,100)$. 
}
\label{f10}
\end{figure}

\begin{figure}[!tbp]
\centering
\subfigure[$t=$12~h]{\includegraphics[width=0.4\textwidth]{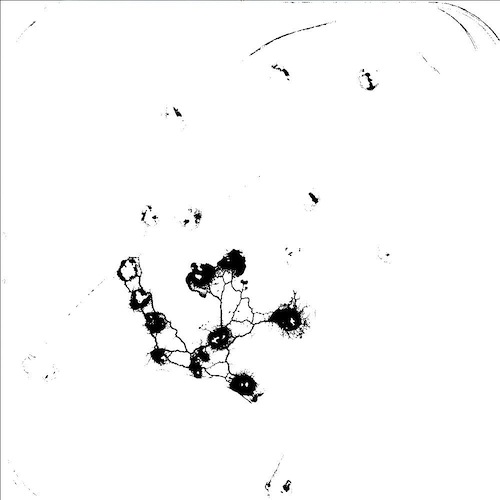}}
\subfigure[$t=$34~h]{\includegraphics[width=0.4\textwidth]{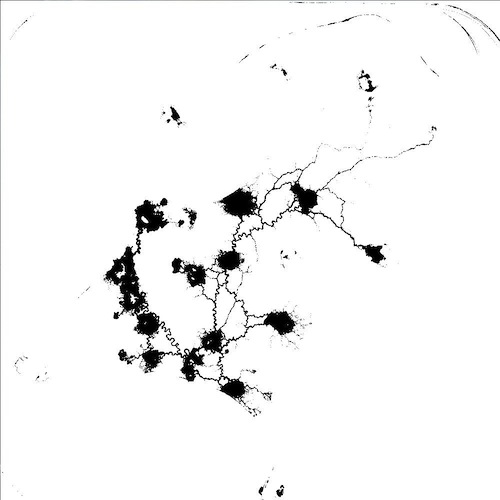}}
\subfigure[$t=$57~h]{\includegraphics[width=0.4\textwidth]{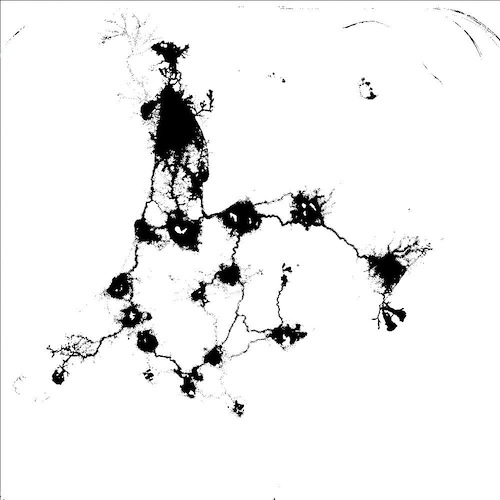}}
\subfigure[$t=$80~h]{\includegraphics[width=0.4\textwidth]{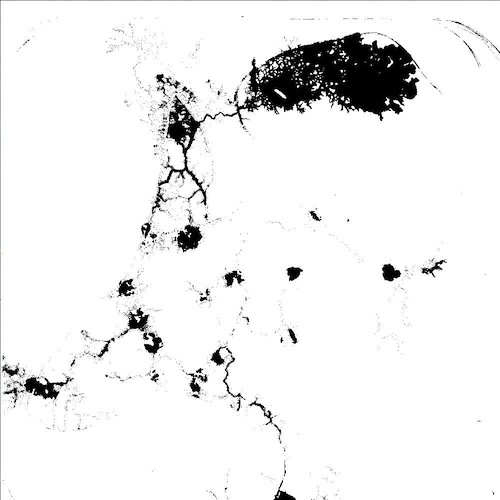}}
\caption{Continued.}
\label{f10}
\end{figure}

The plasmodium of \emph{Physarum polycephalum} rarely repeats itself
in experimental trials. The overall or average pattern, as we will discuss further in
the paper, may be the same but a myriad of variations are possible in the
course of plasmodium's spatial development. 

Outperforming (spreading
out of the dedicated area) and under-performing (not colonising the
whole area) are typical examples of the varieties in plasmodium
behaviour. These two examples are illustrated and discussed below.

In a substantial number of laboratory experiments, the plasmodium did
not stop its foraging activity even when all sources of nutrients were
occupied and the whole agar plate was explored. As shown in
Fig.~\ref{f10} a vigorous plasmodium can spread over surrounding Petri
dishes, trying to settle on bare plastic.

In this experiments Plasmodium starts its colonisation in Amsterdam as
before. It's colinization is more aggresive in this case and it
colonises Haarlem, Den Haag, Rotterdam, Dordrecht, Breda, Tilburg,
Hertogenbosch, Eindhoven within the first twelve hours. A pronounced
protoplasm transport link is established connecting these cities in a
chain (first 12 hours from the moment of inoculation,
Fig.~\ref{f10}ae). 34 hours after inoculation the plasmodium
sprawls from Hertogenbosch to Utrecht, Amersfoort and Apeldoon, and
then builds a transport link Amersfoort-Lelystad-Zwolle-Enschede
(Fig.~\ref{f10}bf).

Protoplasmic tubes connecting Haarlem, Amsterdam, Lelystad with Den
Helder are grown simultaneously after 57 hours of the experiment. By
the same time plasmodium also connects Nijmegen with Apeldoon
(Fig.~\ref{f10}cg). Protoplasmic transport links Den Helder -
Leeuwarden - Groningen, Rotterdam - Middelburg and Eindhoven -
Maastricht are developed by the 80th hour of plasmodium's foraging
activity (Fig.~\ref{f10}dh). A Schematic illustration of the
colonisation dynamics is shown in Fig.~\ref{arrowsf12}b.

Plasmodium starts to show overperformance after 57 hours of the
experiment. It sprawls from Den Helder north-westward and from
Enschede south-eastward onto bare plastic of the experimental
container (Fig.~\ref{f10}cg). The plasmodium does not propagate on the
plastic long enough and retracts in few hours (this can be seen in
Fig.~\ref{f10}dh). Another sprawling takes place by the 80th hour of
experimentation, when plasmodium propagates westward of Middelburg and
sout-eastward of Maasrticht (Fig.~\ref{f10}dh). See also diagrams of
sprawling outside the county in Fig.~\ref{arrowsf12}b. Also notice how
the plasmodium dynamically changes its foraging strategy
(Fig.~\ref{f10}). It first attempts to colonise cities in north-east
part of the country but then abandons the attempt and move to
north-east later via the IJsselmeer lake.

\begin{figure}[!tbp]
\centering
\subfigure[$t=$22~h]{\includegraphics[width=0.4\textwidth]{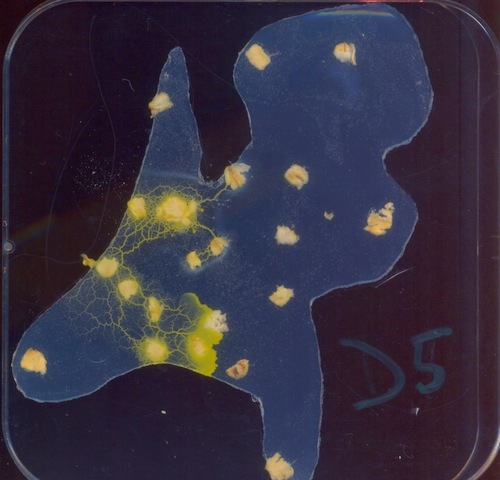}}
\subfigure[$t=$43~h]{\includegraphics[width=0.4\textwidth]{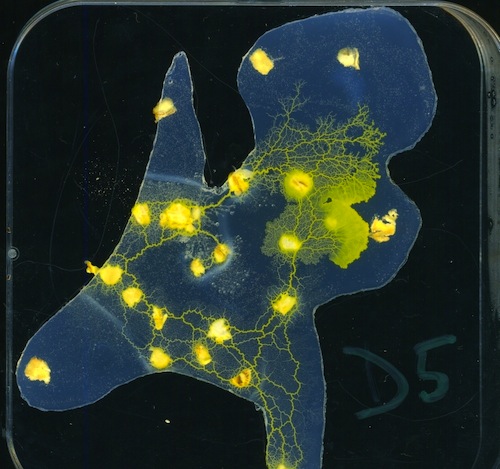}}
\subfigure[$t=$65~h]{\includegraphics[width=0.4\textwidth]{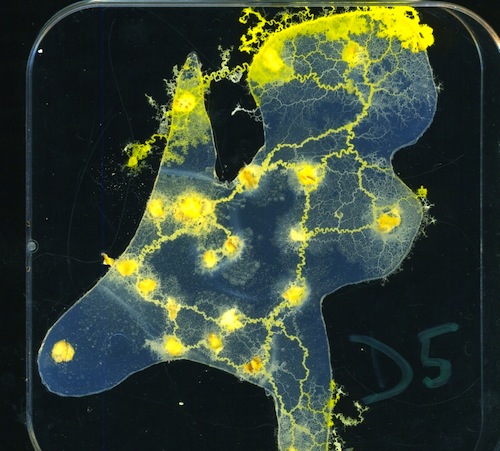}}
\subfigure[$t=$22~h]{\includegraphics[width=0.4\textwidth]{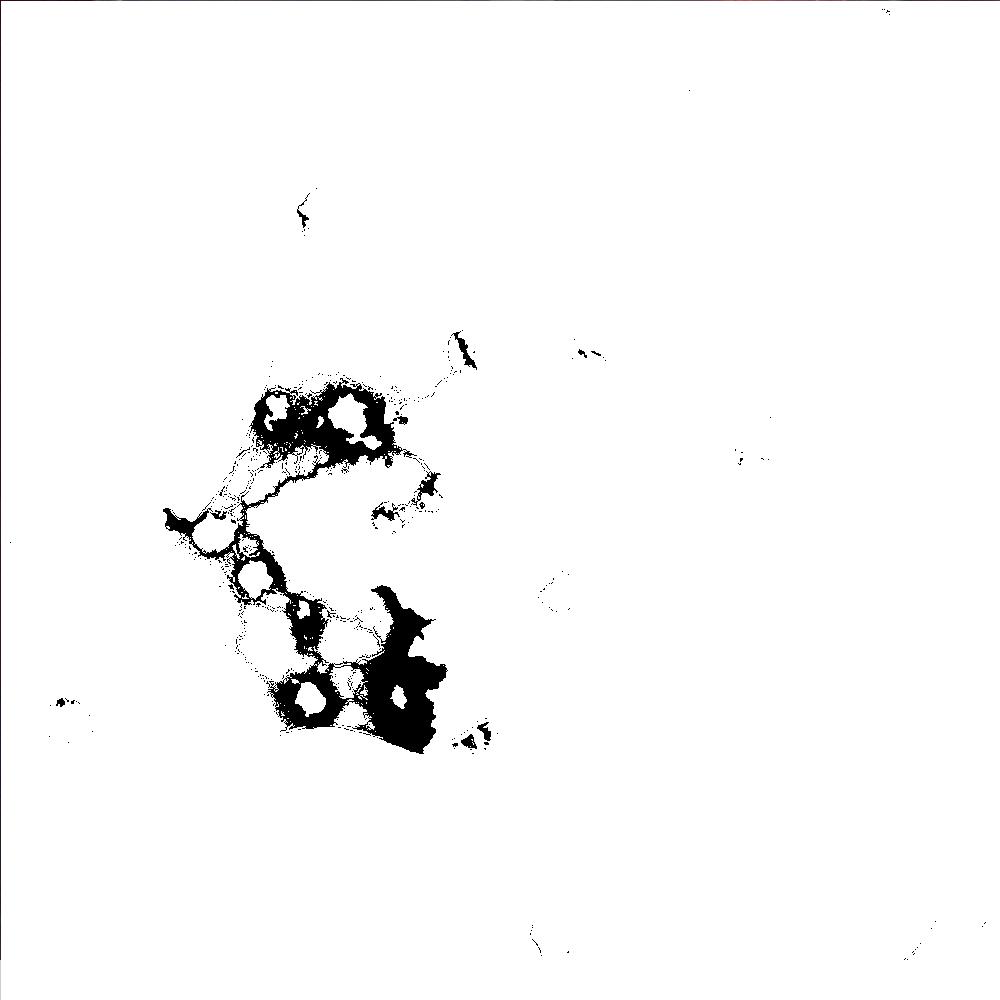}}
\subfigure[$t=$43~h]{\includegraphics[width=0.4\textwidth]{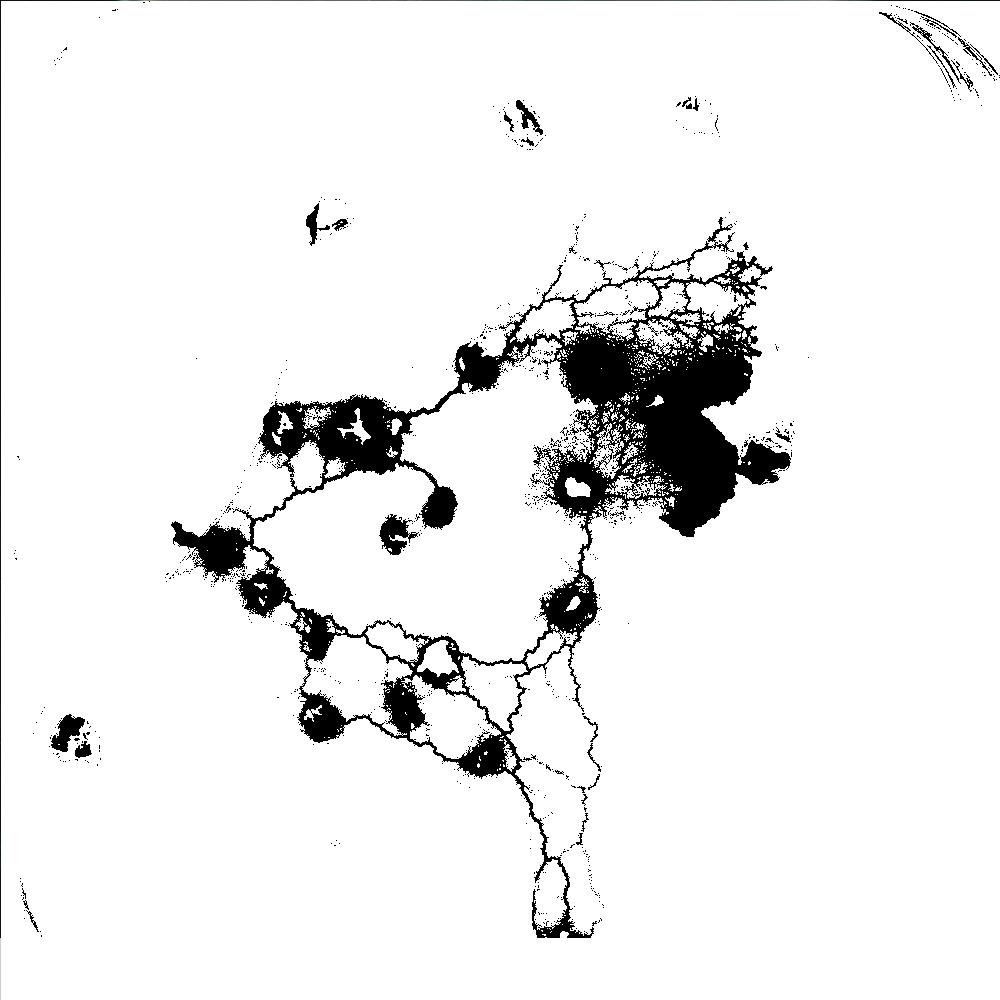}}
\subfigure[$t=$65~h]{\includegraphics[width=0.4\textwidth]{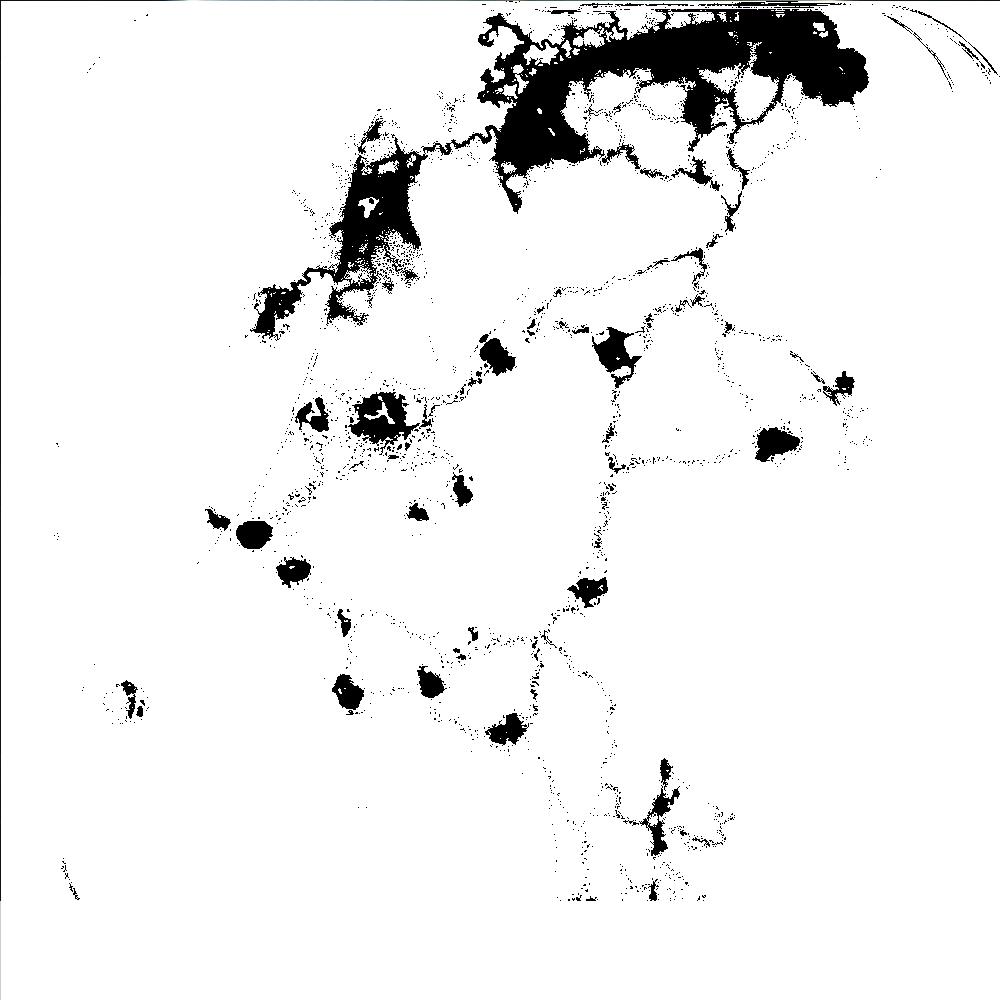}}
\captcont{Plasmodium does not always span all cities (sources of food):
(a)--(d)~scanned image of experimental Petri dish. Time elapsed from inoculation is shown in the sub-figure captions. (e)--(h)~binary images, $\Theta=(100,100,100)$. 
}
\label{fd5}
\end{figure}

In some experiments the plasmodium never manages to span all cities,
and fails to collanize some oat flakes. An example is shown in
(Fig.~\ref{fd5}), 65 hours after inoculation the plasmodium colonises
the majority of the Netherlands and establishes a network of
protoplasmic tubes over most of the cities represented by oat flakes
(Fig.~\ref{fd5}cf). Later it goes into a hibernation stage and forms a
sclerotium. However, at no moment of its development does the
plasmodium even approach Middelburg. Such situations are rather atypical
and did not happen often in our experiments.

As illustrated above, plasmodium is quite an unpredictable creature
and the patterns formed by its protoplasmic networks in any two
experiments rarely match each other exactly. Thus we generalise
results of our experiments by constructing a probabilistic Physarum
graph. A Physarum graph is a tuple ${\mathbf P} = \langle {\mathbf U},
{\mathbf E}, w \rangle$, where $\mathbf U$ is a set of 21 cities,
$\mathbf E$ is a set edges, and $w: {\mathbf E} \rightarrow [0,1]$ is
a probability-weights of edges from $\mathbf E$. For every two cities
$a$ and $b$ from $\mathbf U$ there is an edge connected $a$ and $b$ if
a plasmodium's protoplasmic link is recorded at least in one of $k$
experiments, and the edge $(ab)$ has a probability calculated as a
ratio of experiments where protoplasmic link $(ab)$ occurred to the
total number of experiments $k$. We do not take into account exact
configuration of the protoplasmic tubes but merely their existence,
e.g. protoplasmic tubes linking Eindhoven with Maastricht always
positioned inside the Netherland territory but corresponding edge in
Physarum graph represents the tubes by straight line. We also consider
threshold $\theta \in [0,1]$ Physarum graphs ${\mathbf P}(\theta)$,
defined as follows: for $a,b \in \mathbf U$, $(ab) \in \mathbf E$ if
$w(ab) > \theta$.

\begin{figure}[!tbp]
\centering
\subfigure[$\theta=0$]                {\includegraphics[width=0.24\textwidth] {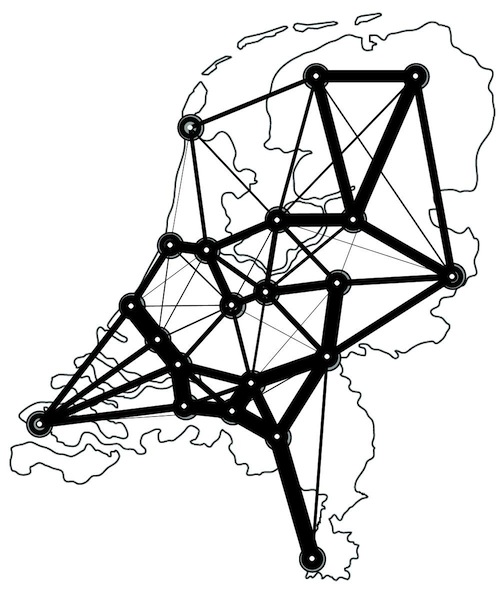}}
\subfigure[$\theta=\frac{1}{62}$]{\includegraphics[width=0.24\textwidth]{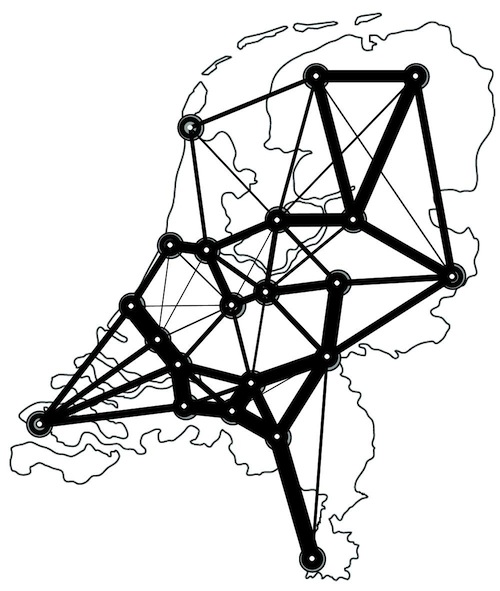}}
\subfigure[$\theta=\frac{2}{62}$]{\includegraphics[width=0.24\textwidth]{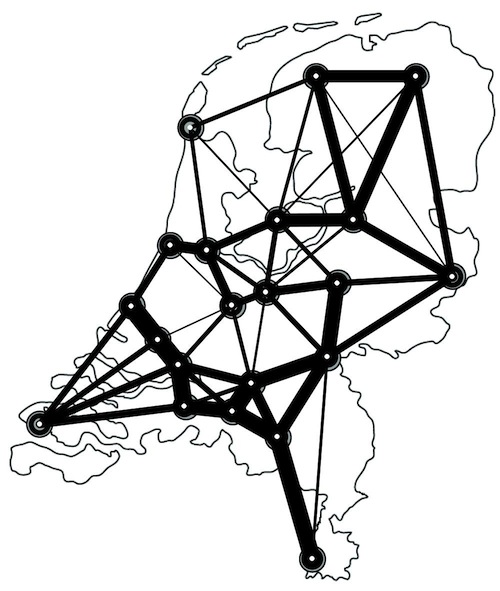}}
\subfigure[$\theta=\frac{3}{62}$]{\includegraphics[width=0.24\textwidth]{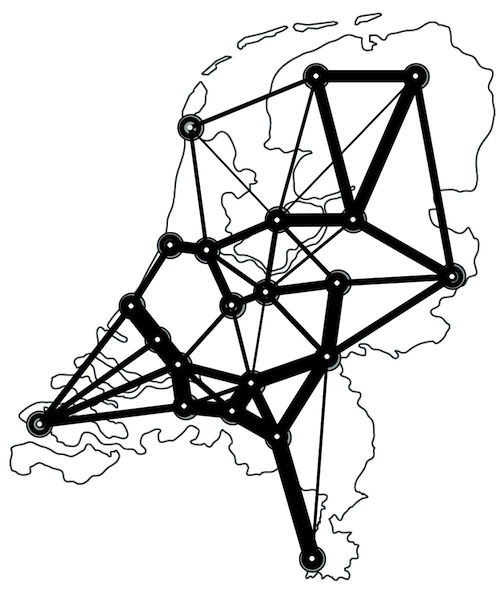}}
\subfigure[$\theta=\frac{4}{62}$]{\includegraphics[width=0.24\textwidth]{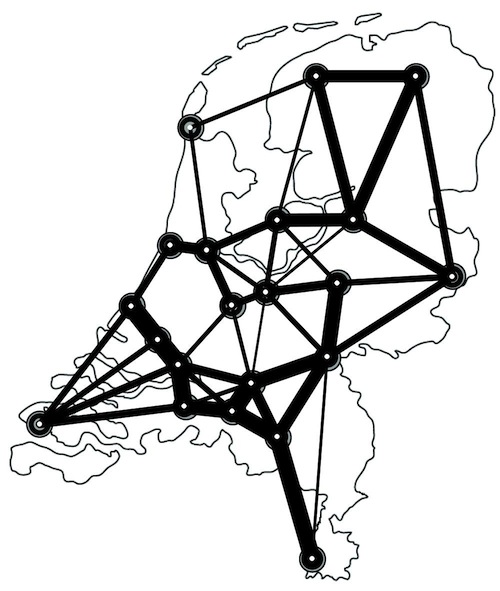}}
\subfigure[$\theta=\frac{5}{62}$]{\includegraphics[width=0.24\textwidth]{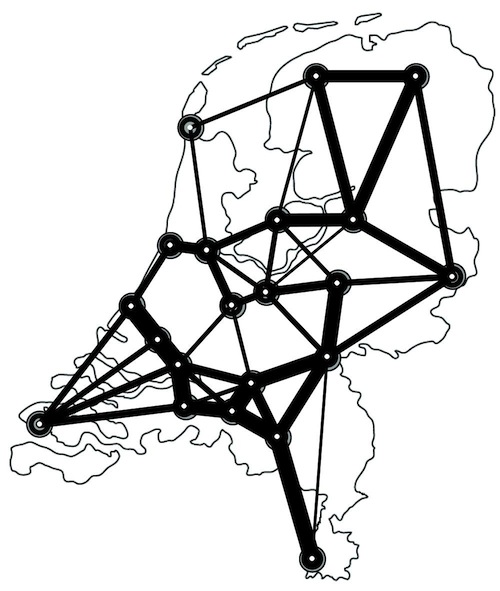}}
\subfigure[$\theta=\frac{6}{62}$]{\includegraphics[width=0.24\textwidth]{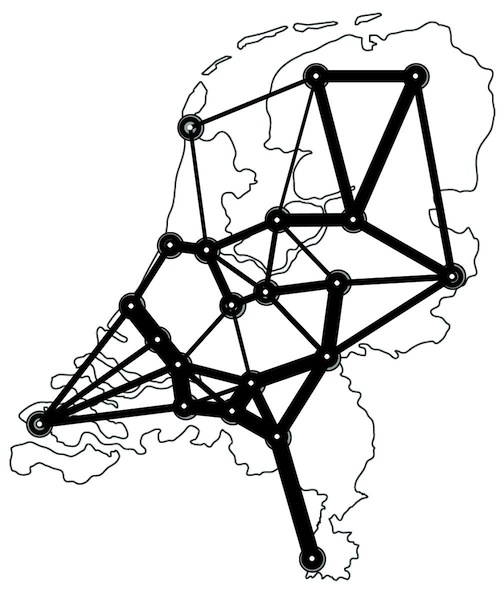}}
\subfigure[$\theta=\frac{7}{62}$]{\includegraphics[width=0.24\textwidth]{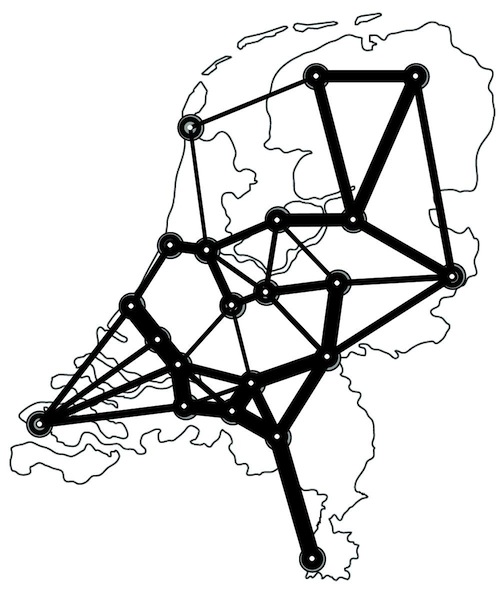}}
\subfigure[$\theta=\frac{8}{62}$]{\includegraphics[width=0.24\textwidth]{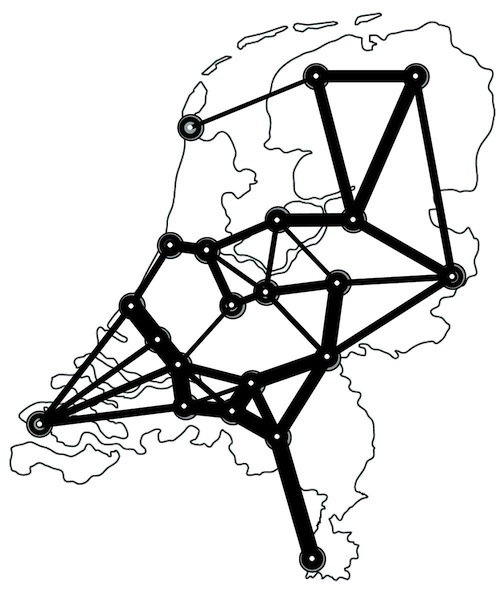}}
\subfigure[$\theta=\frac{9}{62}$]{\includegraphics[width=0.24\textwidth]{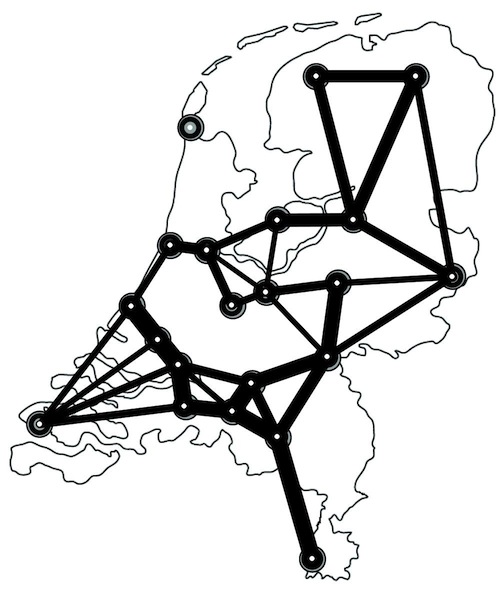}}
\subfigure[$\theta=\frac{10}{62}$]{\includegraphics[width=0.24\textwidth]{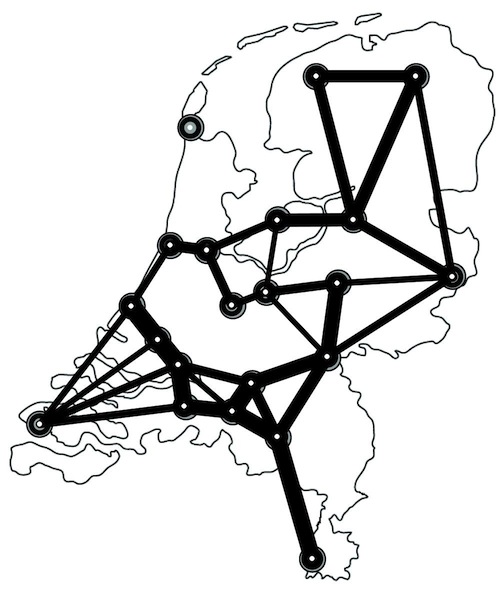}}
\subfigure[$\theta=\frac{11}{62}$]{\includegraphics[width=0.24\textwidth]{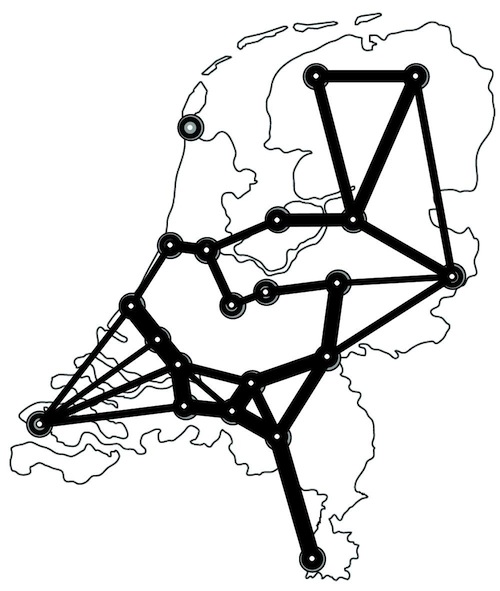}}
\subfigure[$\theta=\frac{12}{62}$]{\includegraphics[width=0.24\textwidth]{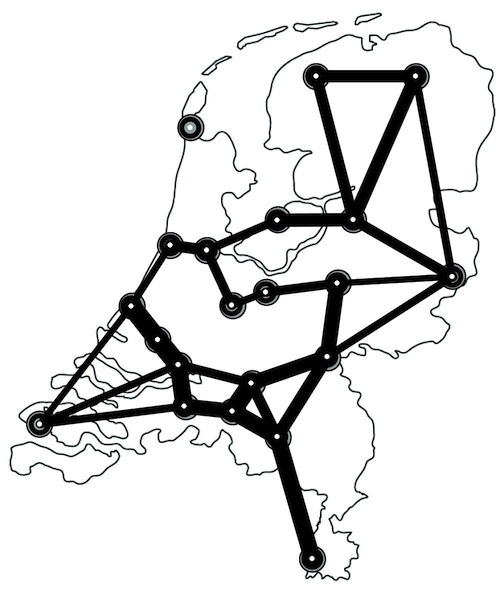}}
\subfigure[$\theta=\frac{13}{62}$]{\includegraphics[width=0.24\textwidth]{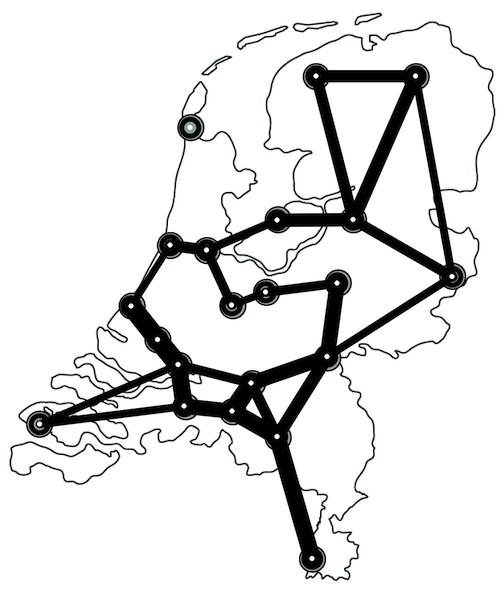}}
\subfigure[$\theta=\frac{14}{62}$]{\includegraphics[width=0.24\textwidth]{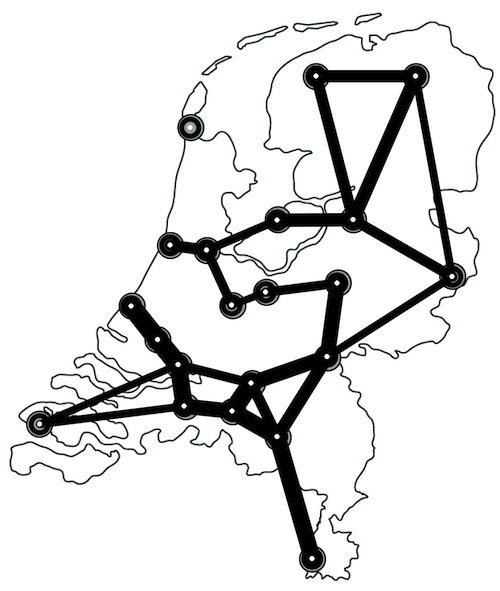}}
\subfigure[$\theta=\frac{15}{62}$]{\includegraphics[width=0.24\textwidth]{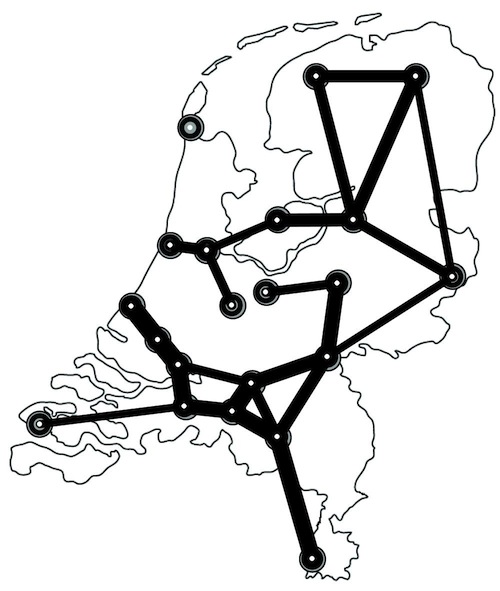}}
\caption{Configurations of threshold \emph{Physarum}-graph ${\mathbf P}(\theta)$ 
for $\theta=0, \frac{1}{62}, \ldots, \frac{15}{62}$ Thickness of an edge is proportional to the edge's weight.}
\label{Fgraphs00_15}
\end{figure}

\begin{figure}[!tbp]
\centering
%\subfigure[$\theta=0$]{\includegraphics[width=0.24\textwidth]{figs/PhysarumGraph/physarum_00}}
\subfigure[$\theta=\frac{16}{62}$]{\includegraphics[width=0.24\textwidth]{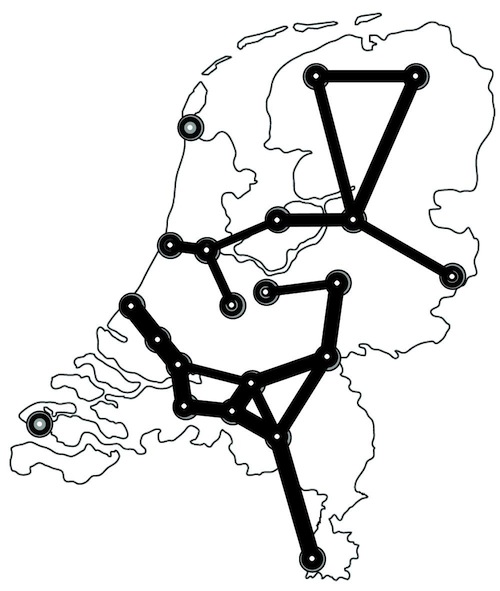}}
\subfigure[$\theta=\frac{17}{62}$]{\includegraphics[width=0.24\textwidth]{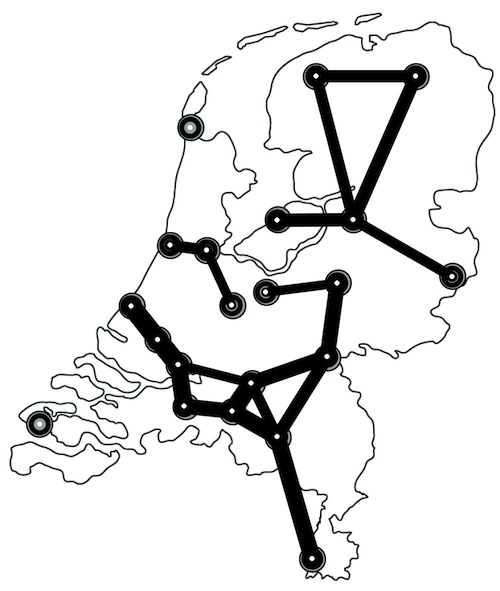}}
\subfigure[$\theta=\frac{19}{62}$]{\includegraphics[width=0.24\textwidth]{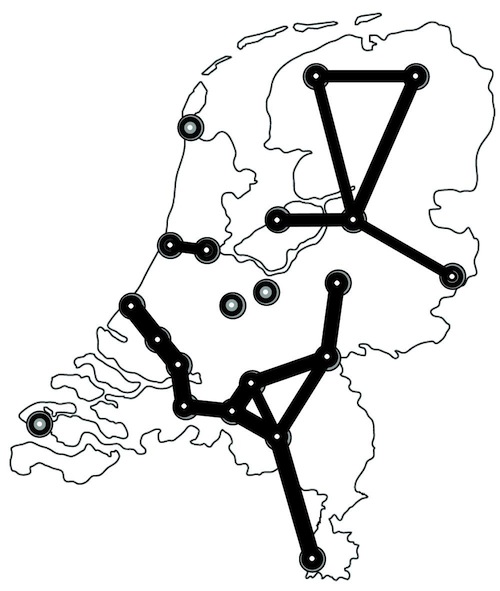}}
\subfigure[$\theta=\frac{21}{62}$]{\includegraphics[width=0.24\textwidth]{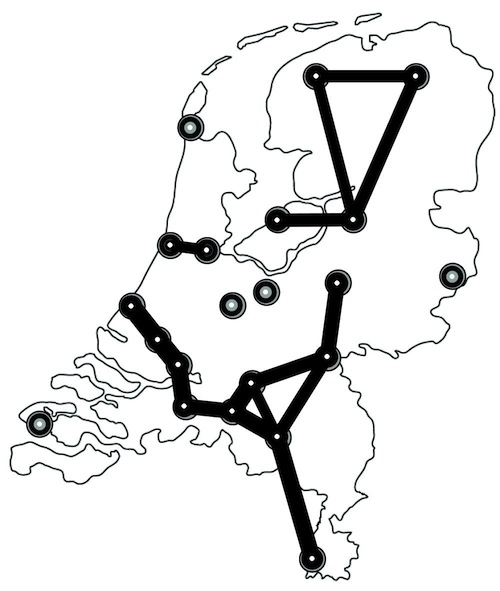}}
\subfigure[$\theta=\frac{22}{62}$]{\includegraphics[width=0.24\textwidth]{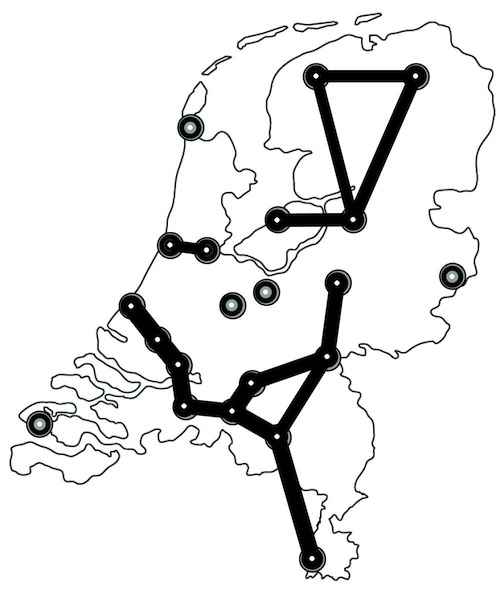}}
\subfigure[$\theta=\frac{23}{62}$]{\includegraphics[width=0.24\textwidth]{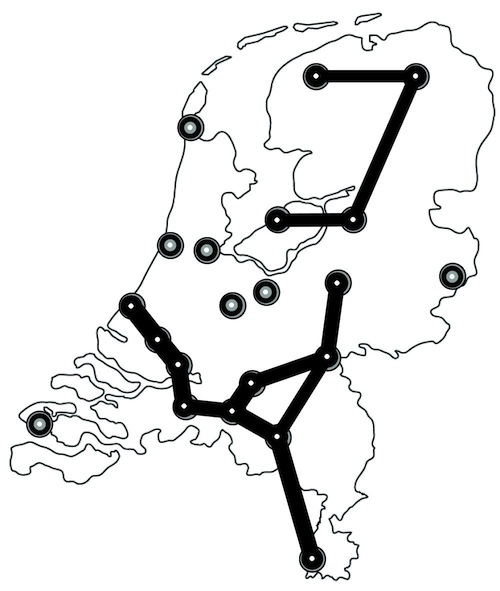}}
\subfigure[$\theta=\frac{25}{62}$]{\includegraphics[width=0.24\textwidth]{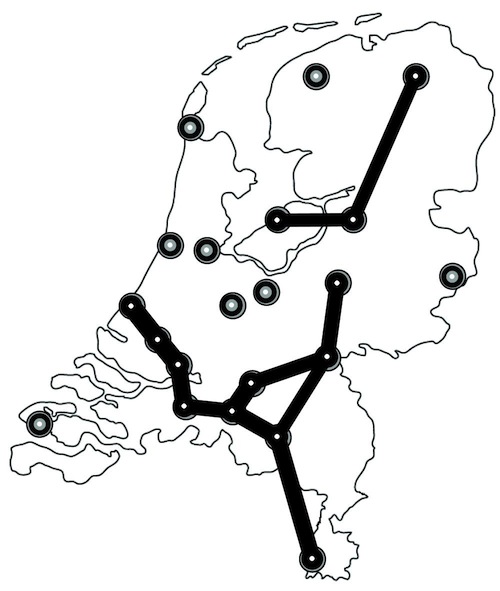}}
\subfigure[$\theta=\frac{26}{62}$]{\includegraphics[width=0.24\textwidth]{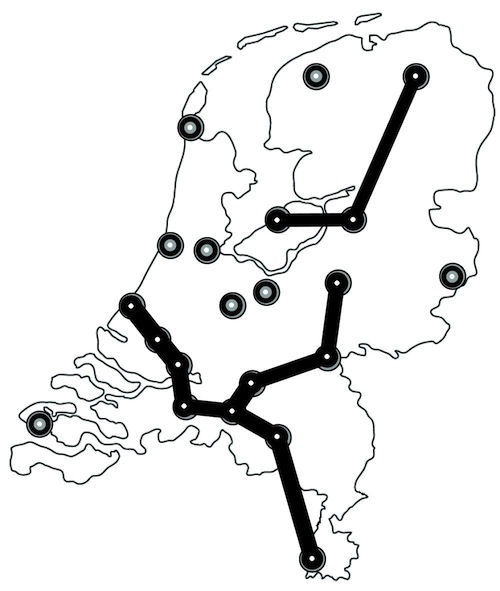}}
\subfigure[$\theta=\frac{27}{62}$]{\includegraphics[width=0.24\textwidth]{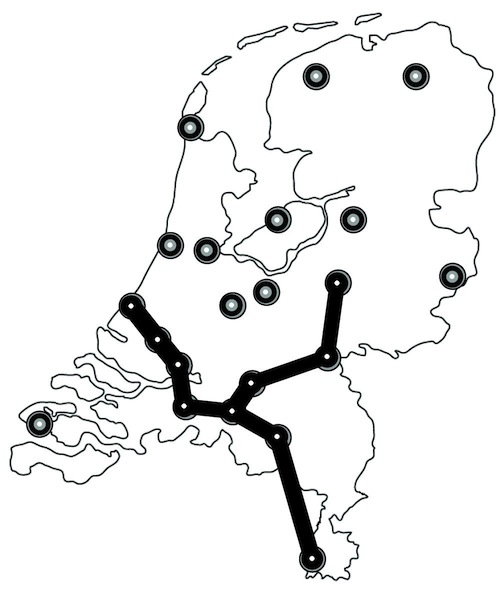}}
\subfigure[$\theta=\frac{30}{62}$]{\includegraphics[width=0.24\textwidth]{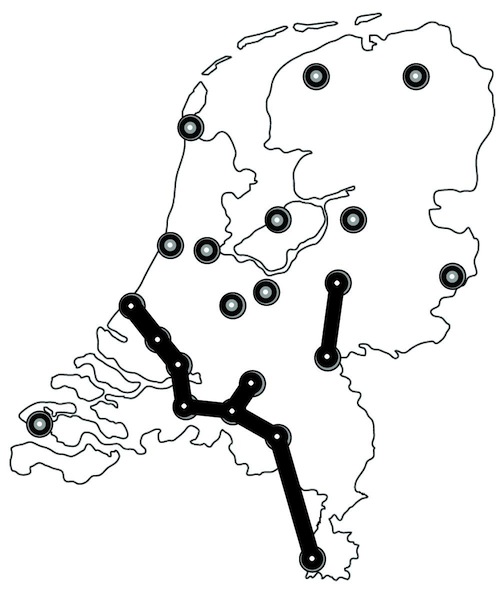}}
\subfigure[$\theta=\frac{31}{62}$]{\includegraphics[width=0.24\textwidth]{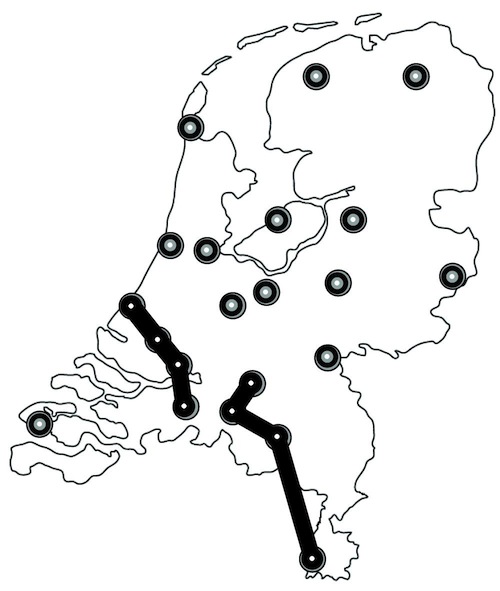}}
\subfigure[$\theta=\frac{34}{62}$]{\includegraphics[width=0.24\textwidth]{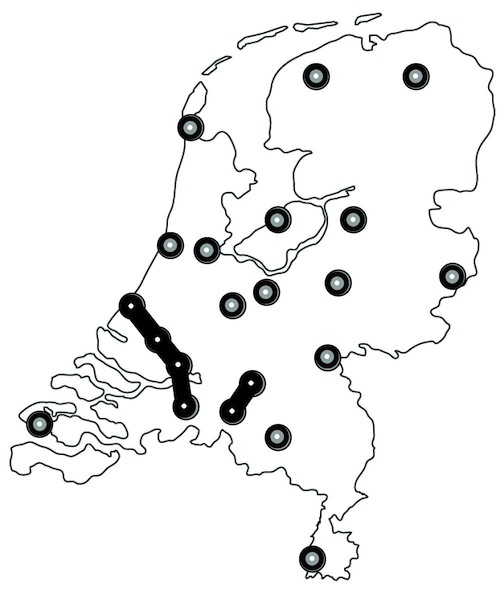}}
\subfigure[$\theta=\frac{35}{62}$]{\includegraphics[width=0.24\textwidth]{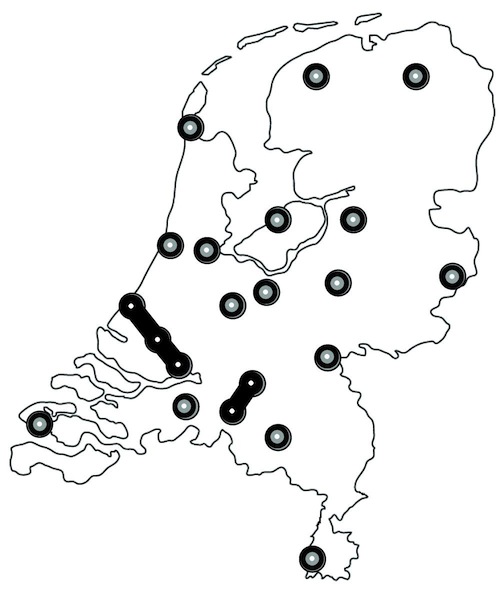}}
\subfigure[$\theta=\frac{37}{62}$]{\includegraphics[width=0.24\textwidth]{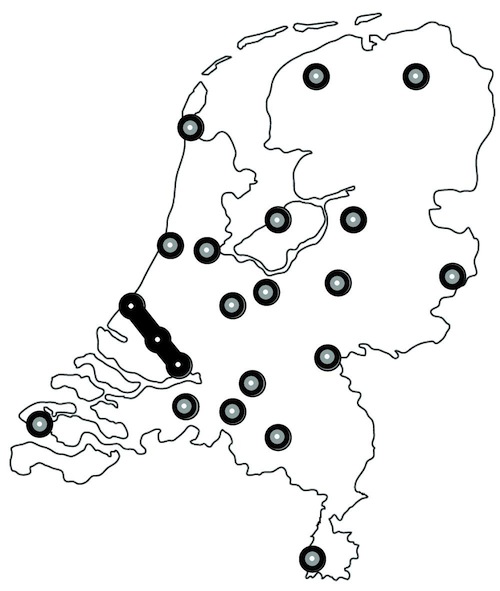}}
%\subfigure[$\theta=\frac{38}{62}$]{\includegraphics[width=0.24\textwidth]{figs/PhysarumGraph/physarum_38}}
\caption{Configurations of threshold \emph{Physarum}-graph ${\mathbf P}(\theta)$ 
for $\theta=\frac{16}{62}, \frac{17}{62}, \ldots, \frac{37}{62}$ Thickness of an edge is proportional to the edge's weight.}
\label{Fgraphs16_38}
\end{figure}

Threshold Physarum-graphs extracted from 62 laboratory experiments are
shown in Figs.~\ref{Fgraphs00_15} and~\ref{Fgraphs16_38}. The graph
becomes planar only for $\theta=\frac{9}{62}$
(Figs.~\ref{Fgraphs00_15}p), i.e. when edges occurred in over 15\% of the
experiments. We can therefore infer that the Physarum graph is
planar. However, with acquiring planarity the graph becomes
disconnected: one node, Den Helder city, becomes isolated.

The Physarum-graphs become acyclic for $\theta=\frac{26}{62}$
(Fig.~\ref{Fgraphs16_38}h), i.e., when its edges appear as protoplasmic
tubes in over 41\% of the experiments. When the graph becomes acyclic it
is split into a set of isolated nodes: Den Helder, Leeuwarden,
Haarlem, Amsterdam, Utrecht, Amersfoort, Enschede, Middelburg, and
two additional components. One component is a chain of three cities: Lelystad,
Zwolle and Groningen. The second component is a tree routed in Tilburg. The
tree has three linear branches:
\begin{itemize}
\item Tilburg - Breda - Dordrecht - Rotterdam - Den Haag
\item Tilburg - Hertogenbosch - Nijmegen - Apeldoon
\item Tilburg - Eindhoven - Maastricht.
\end{itemize}
This tree is a characteristic feature of the Physarum-graph and it
appears in over 60\% of experiments. The tree is `destroyed' when
$\theta \geq \frac{30}{62}$ (Fig.~\ref{Fgraphs16_38}j), then only
chains remain, which give away isolated nodes with further increase of
$\theta$. Some chains are more stable then others. Thus, the chain
Breda - Dordrecht - Rotterdam - Den Haag appears in over 54\% of
experiments (Fig.~\ref{Fgraphs16_38}l). While the chain Dordrecht -
Rotterdam - Den Haag appears in almost 60\% of experiments.

\begin{figure}[!tbp]
\centering
\subfigure[${\mathbf H}$]{\includegraphics[width=0.49\textwidth]{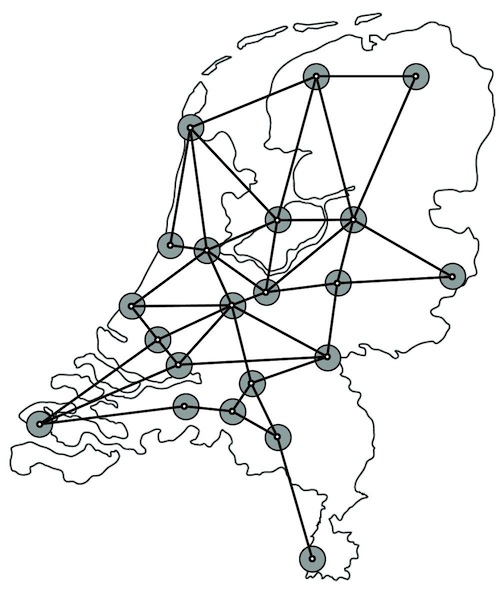}}
\subfigure[${\mathbf P}(0) \bigcap {\mathbf H}$]{\includegraphics[width=0.49\textwidth]{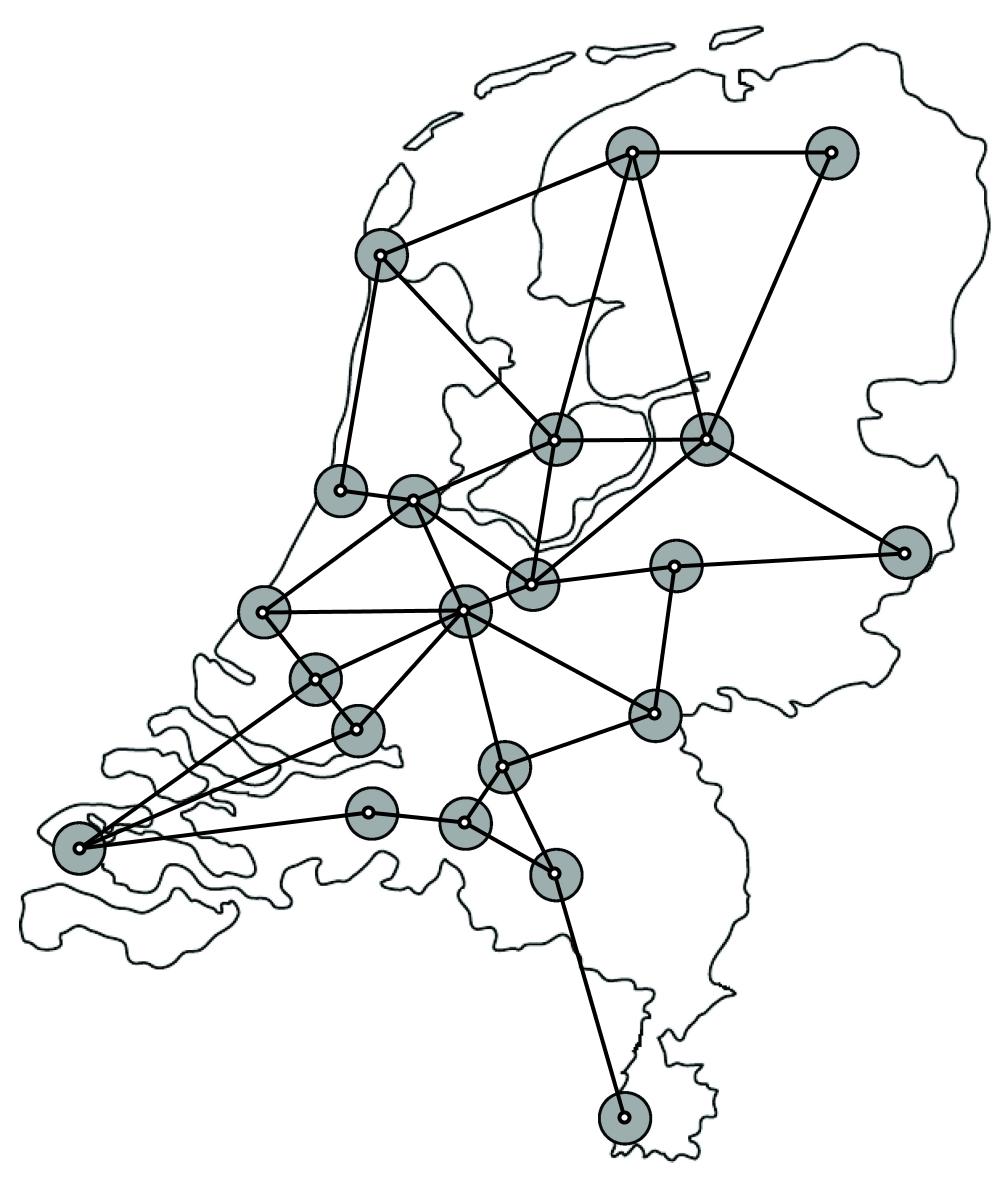}}
\subfigure[${\mathbf P}(\frac{8}{62}) \bigcap {\mathbf H}$]{\includegraphics[width=0.49\textwidth]{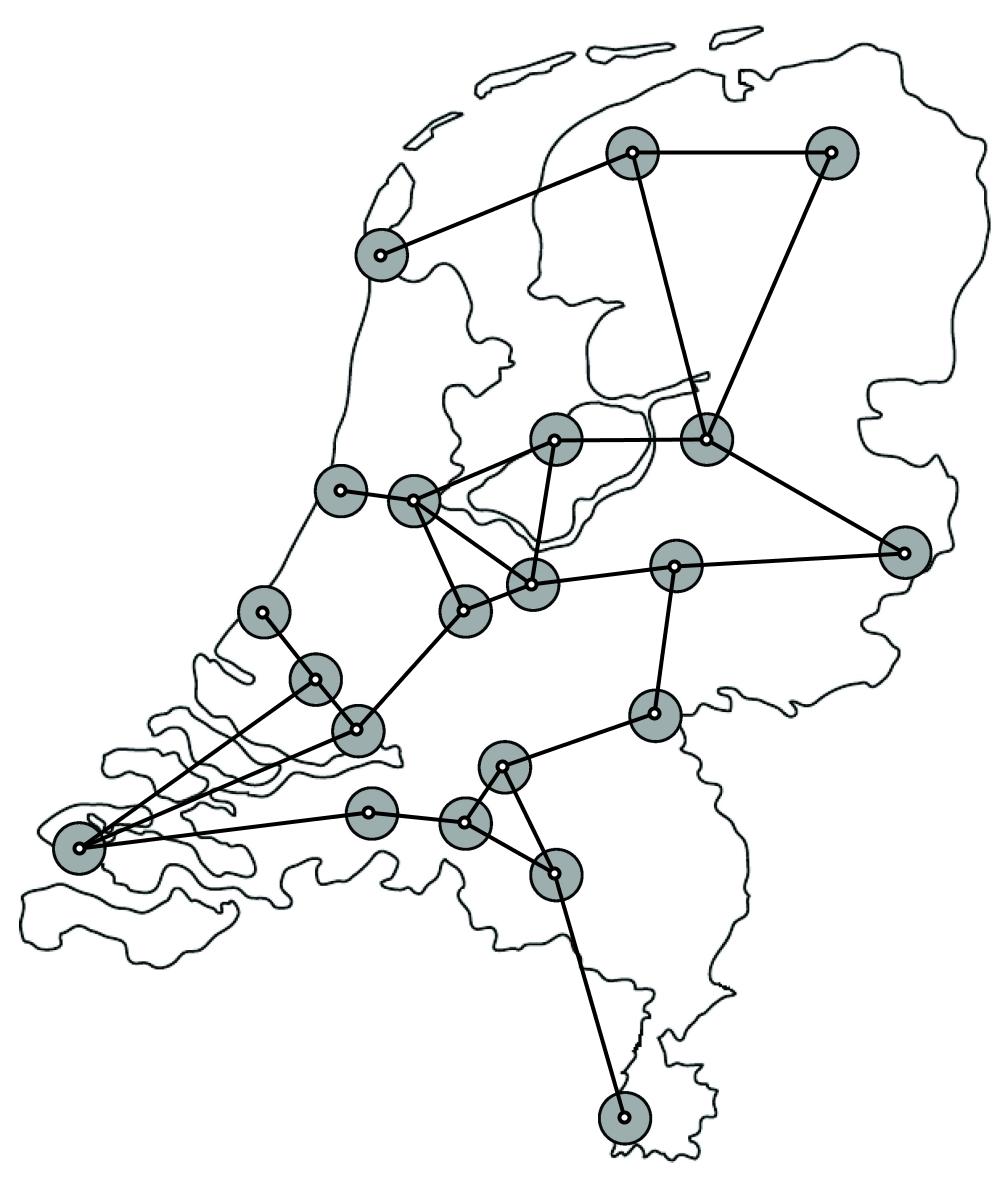}}
\subfigure[${\mathbf P}(\frac{15}{62}) \bigcap {\mathbf H}$]{\includegraphics[width=0.49\textwidth]{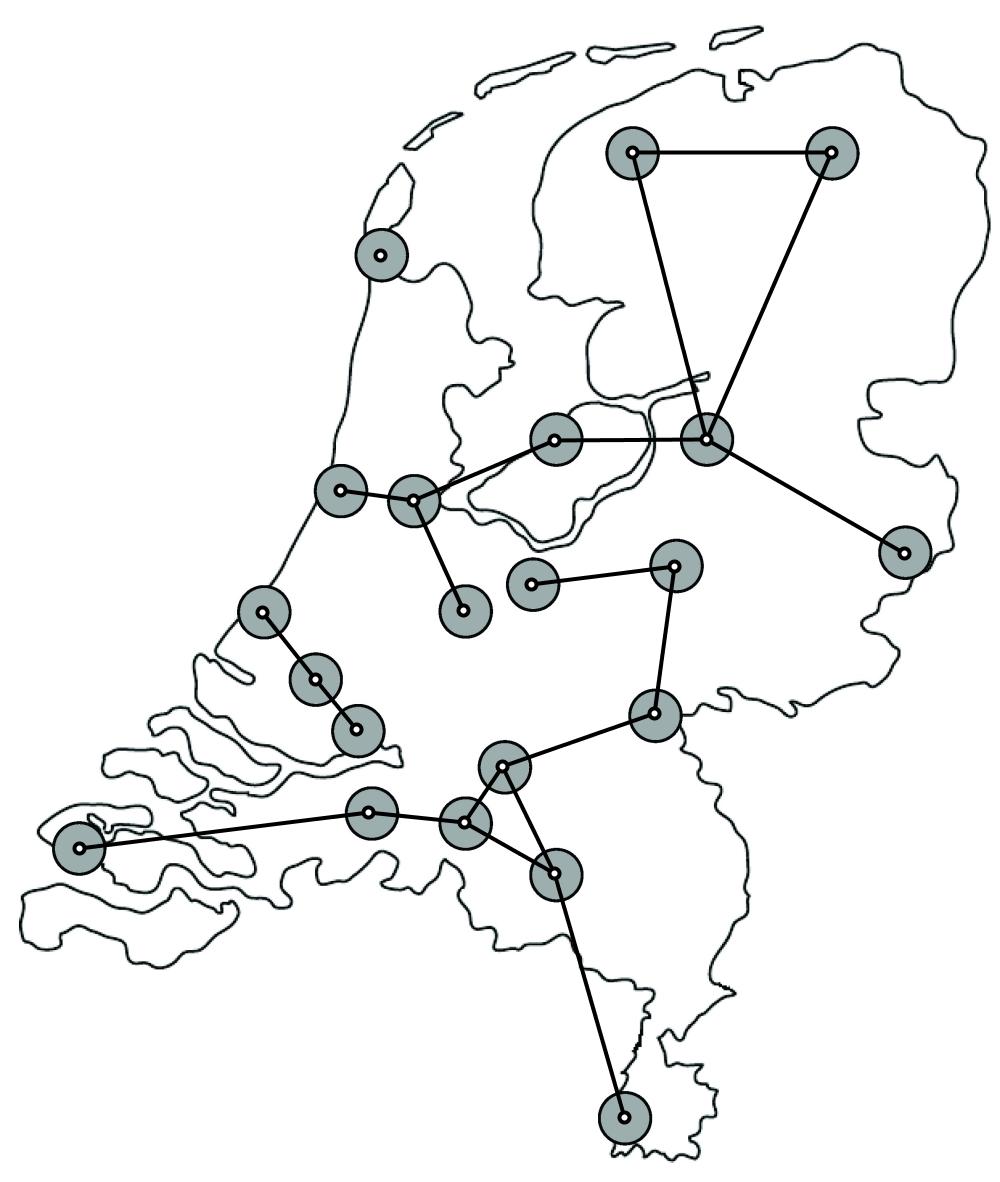}}
\caption{Graph $\mathbf H$ of man-made motorway network is shown in (a). Intersection ${\mathbf P}(\theta) \bigcap {\mathbf H}$,
$\theta=1$, $\frac{8}{16}$ and $\frac{15}{16}$, of Physarum $\mathbf P$ and motorways $\mathbf H$ graphs is shown in (bcd).}
\label{motorways}
\end{figure}

The experiments have now provided a reasonably consistent set of
connections between the various urban centres in the Netherlands. The
next question is to asses how well these Physarum graphs approximate
the Netherlands motorway network. A graph $\mathbf H$ of Dutch
motorways is Fig.~\ref{motorways}a. We construct the motorway graph
$\mathbf H$ as follows. Let $\mathbf U$ be a set of urban
regions/cities, for any two regions $a$ and $b$ from $\mathbf U$, the
nodes $a$ and $b$ are connected by an edge $(ab)$ if there is a
motorway starting in vicinity of $a$ and passing in vicinity of $b$
and not passing in vicinity of any other urban area $c \in \mathbf
U$. If there is a branching motorways, which e.g. starts in $a$ goes
in the direction of $b$ and $c$ and at some point branches towards $b$
and $c$, we then add two separate edges $(ab)$ and $(ac)$ to the graph
$\mathbf H$.
 
The intersection ${\mathbf P}(\theta) \bigcap {\mathbf H}$ of Physarum and
motorways graphs is shown in Fig.~\ref{motorways}bcd for $\theta=0$,
$\frac{8}{16}$ and $\frac{15}{16}$. A relaxed probabilistic Physarum
graph ${\mathbf P}(0)$, where an edge appears in the graph if it is
recorded in at least one experiment, matches the motorway graph
$\mathbf H$ almost perfectly. Just three edges of $\mathbf H$ are not
presented in ${\mathbf P}(0) \bigcap {\mathbf H}$:
\begin{itemize}
\item (Amsterdam, Der Helden), 
\item (Zwolle, Apeldoon),
\item (Rotterdam, Dordrecht) (Fig.~\ref{motorways}b).
\end{itemize}  
$\theta=\frac{8}{16}$ is the highest value for which Physarum graph ${\mathbf P}(\theta)$ remains connected 
(Fig.~\ref{Fgraphs00_15}). Graph  ${\mathbf P}(\frac{8}{62}) \bigcap {\mathbf H}$ loses few more edges 
(presented in ${\mathbf P}(\frac{0}{62}) \bigcap {\mathbf H}$: 
\begin{itemize}
\item (Den Helder, Lelystad), 
\item (Den Helder, Haarlem),
\item (Leeuwarden, Lelystad),
\item (Utrecht, Dordrecht),
\item (Utrecht, Nijmegen),
\item (Breda, Enschede), 
\item (Utrecht, Enschede) (Fig.~\ref{motorways}c).
\end{itemize} 
As soon as $\theta$ reaches value $\frac{15}{16}$ the graph ${\mathbf P}(\theta) \bigcap {\mathbf H}$ becomes separated
on isolated node Den Helder, and there components: 
\begin{itemize}
\item chain Enschede - Den Haag -Rotterdam,
\item cycle Nijmegen - Breda - Middelburg with branches Breda - Hertogenbosch - Tilburg, Middelburg - Eindhoven and 
Nijmegen - Dordrecht - Apeldoon - Amersfoort,
\item cycle Leeuwarden - Groningen - Zwolle with branches Zwolle - Enschede and 
Zwolle - Lelystad - Amsterdam - {Haarlem, Utrecht} (Fig.~\ref{motorways}d). 
\end{itemize}

\section{Comparing motorway and Physarum graphs with proximity graphs}
\label{proximitygraphs}

We conjectured that plasmodium of \emph{Physarum polycephalum}
constructs planar proximity graphs by its protoplasmic
network~\cite{adamatzky_ppl_2009}. A protoplasmic network constructed
in any particular experiment is planar, a generalised Physarum graph
$\mathbf P$ may be non-planar. A planar graph consists of nodes which
are points on a Euclidean plane with edges which are straight segments
connecting the points.  A planar proximity graph is a planar graph
where two points are connected by an edge if they are close in some
sense. Usually a pair of points are assigned a certain neighborhood,
and the pair are connected by an edge if their neighborhood is empty.
Relative neighborhood graph~\cite{jaromczyk_toussaint_1992}, Gabriel
graph~\cite{matula_sokal_1984}, $\beta$-skeletons~\cite{kirkpatrick}
and spanning tree are most known examples of proximity graphs.

\begin{figure}[!tbp]
\centering
\subfigure[$\mathbf{RNG}$]{\includegraphics[width=0.49\textwidth]{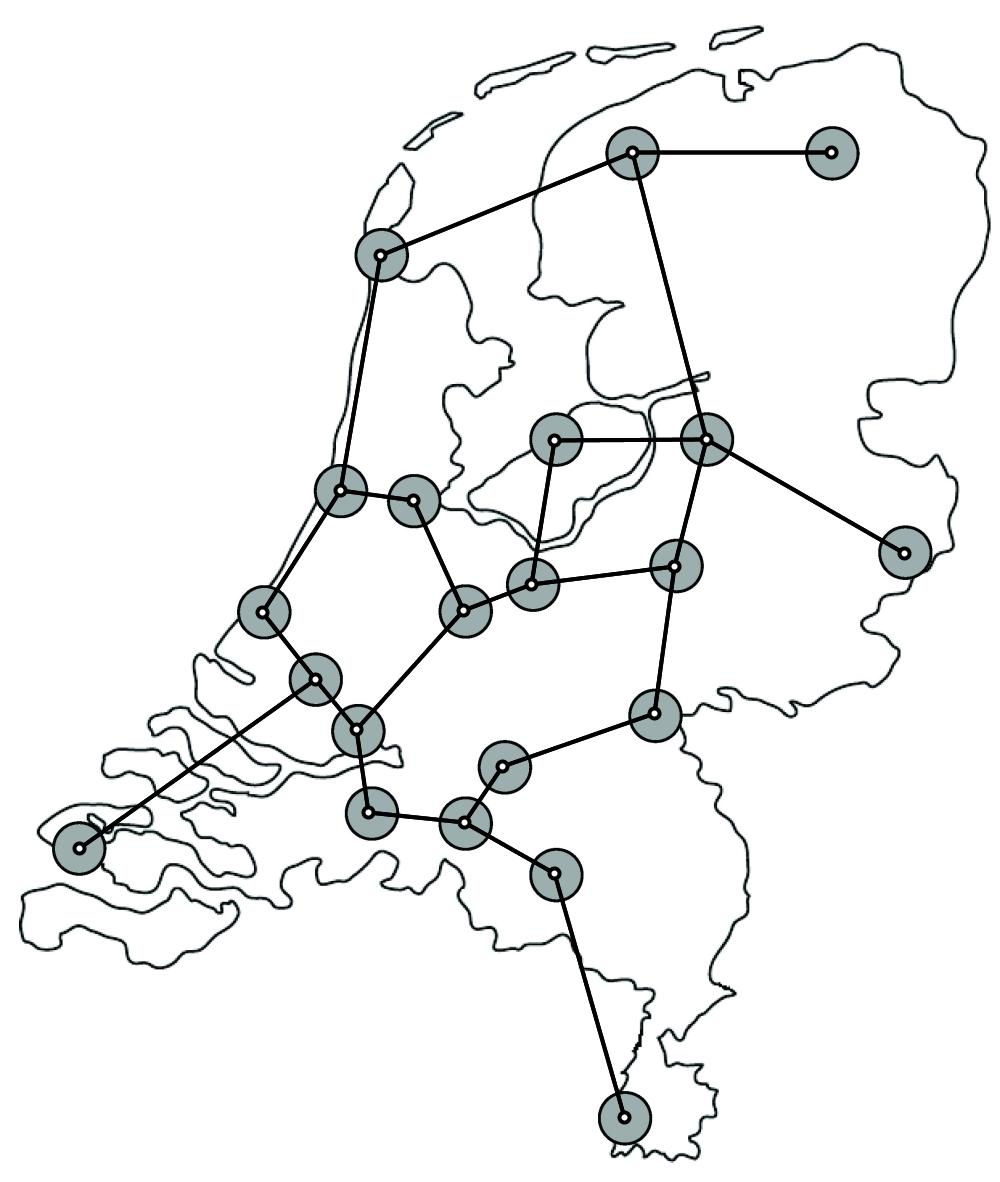}}
\subfigure[$\mathbf{BS}$(1.5)]{\includegraphics[width=0.49\textwidth]{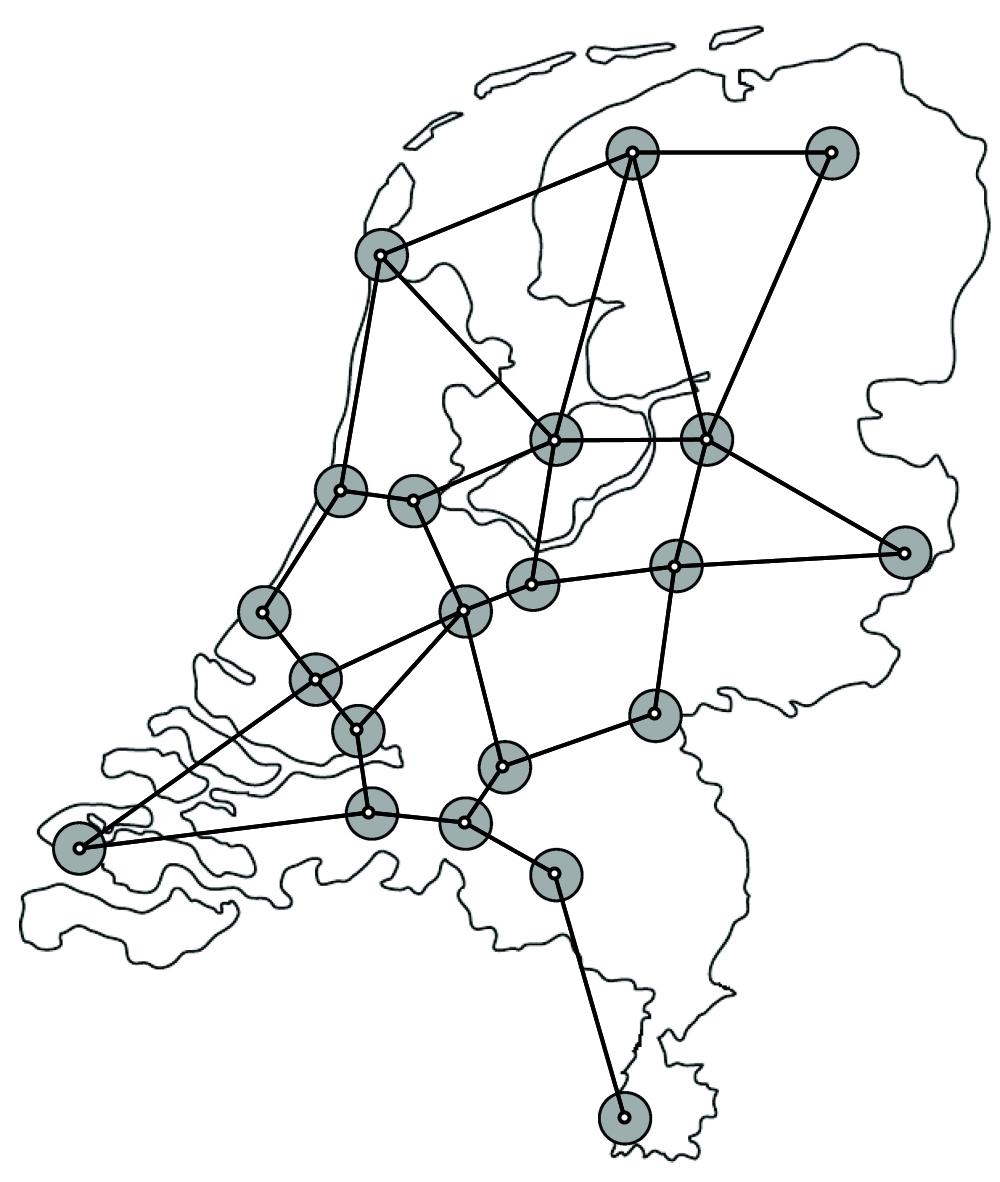}}
\subfigure[$\mathbf{GG}$]{\includegraphics[width=0.49\textwidth]{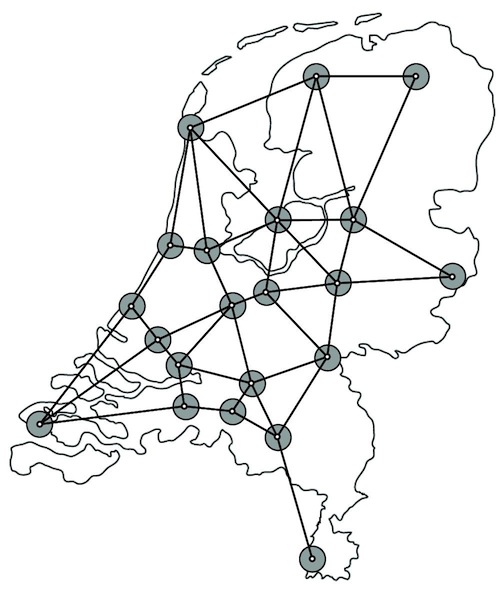}}
\subfigure[$\mathbf{MST}$]{\includegraphics[width=0.49\textwidth]{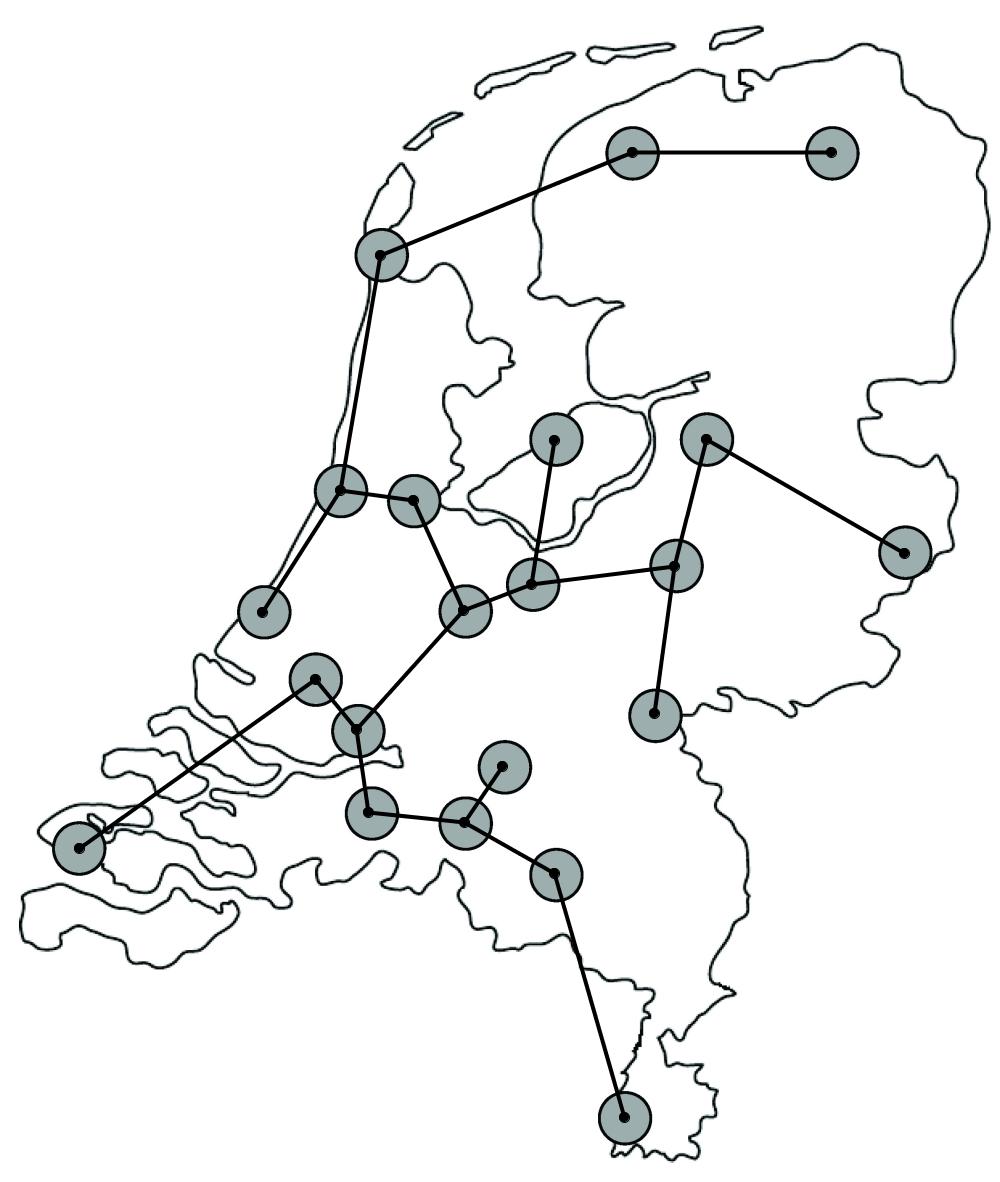}}
\caption{Proximity graphs constructed on regions $\mathbf U$: 
(a)~relative neighbourhood graph $\mathbf{RNG}$,
(b)~$\beta$-skeleton with control parameter 1.5, $\mathbf{BS}(1.5)$
(c)~Gabriel graph $\mathbf{GG}$,
(d)~Minimum spanning tree $\mathbf{MST}$}
\label{proximity}
\end{figure}

For self-consistency we provide a brief definition of the following graphs:
\begin{itemize}
\item $\mathbf{RNG}$: Points $a$ and $b$ are connected by an edge in
  $\mathbf{RNG}$ if no other point $c$ is closer to $a$ and $b$ than
  $dist(a,b)$~\cite{toussaint_1980}.
\item $\mathbf{GG}$: Points $a$ and $b$ are connected by an edge in
  $\mathbf{GG}$ if a disc with diameter $dist(a,b)$ centered in middle
  of the segment $ab$ is empty~\cite{gabriel_sokal_1969,matula_sokal_1984}.
\item $\mathbf{BS}(\beta)$: A $\beta$-skeleton, $\beta \geq 1$, is a
  planar proximity undirected graph of an Euclidean point set where
  nodes are connected by an edge if their lune-based neighborhood
  contains no other points of the given set; parameter $\beta$
  determines size and shape of the nodes'
  neighborhoods~\cite{kirkpatrick}.
\item $\mathbf{MST}$: The Euclidean minimal spanning tree
  (MST)~\cite{nesetril} is a connected acyclic graph which has minimal
  possible sum of edges' lengths.
\end{itemize} 
The graphs are related as $\mathbf{MST} \subseteq \mathbf{RNG}
\subseteq \mathbf{BS}
\subseteq\mathbf{GG}$~\cite{toussaint_1980,matula_sokal_1984,jaromczyk_toussaint_1992}.

\begin{figure}[!tbp]
\centering
\subfigure[$\mathbf P \bigcap \mathbf{ RNG}$ ]{\includegraphics[width=0.32\textwidth]{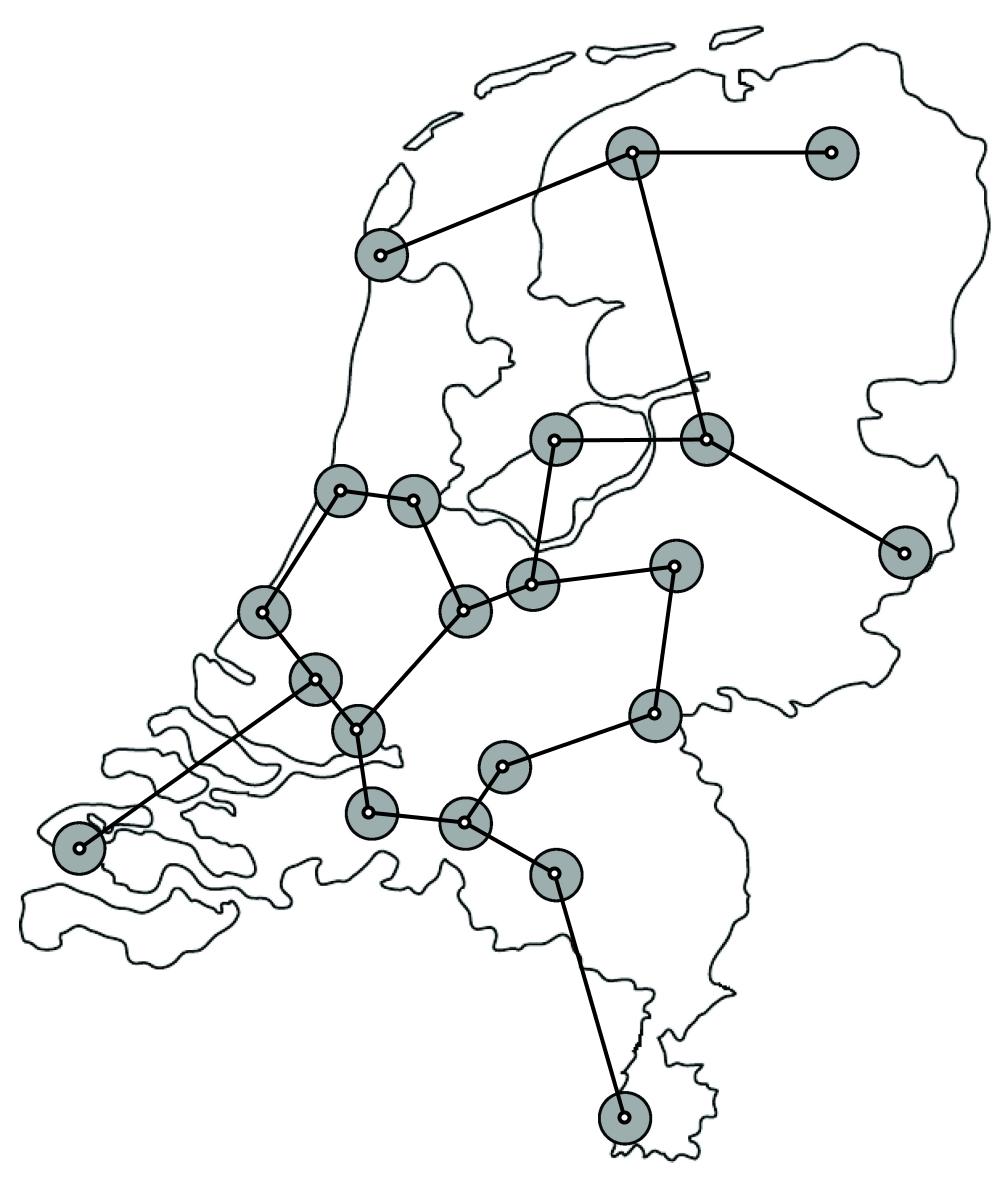}}
\subfigure[$\mathbf P \bigcap \mathbf{ BS}$(1.5)]{\includegraphics[width=0.32\textwidth]{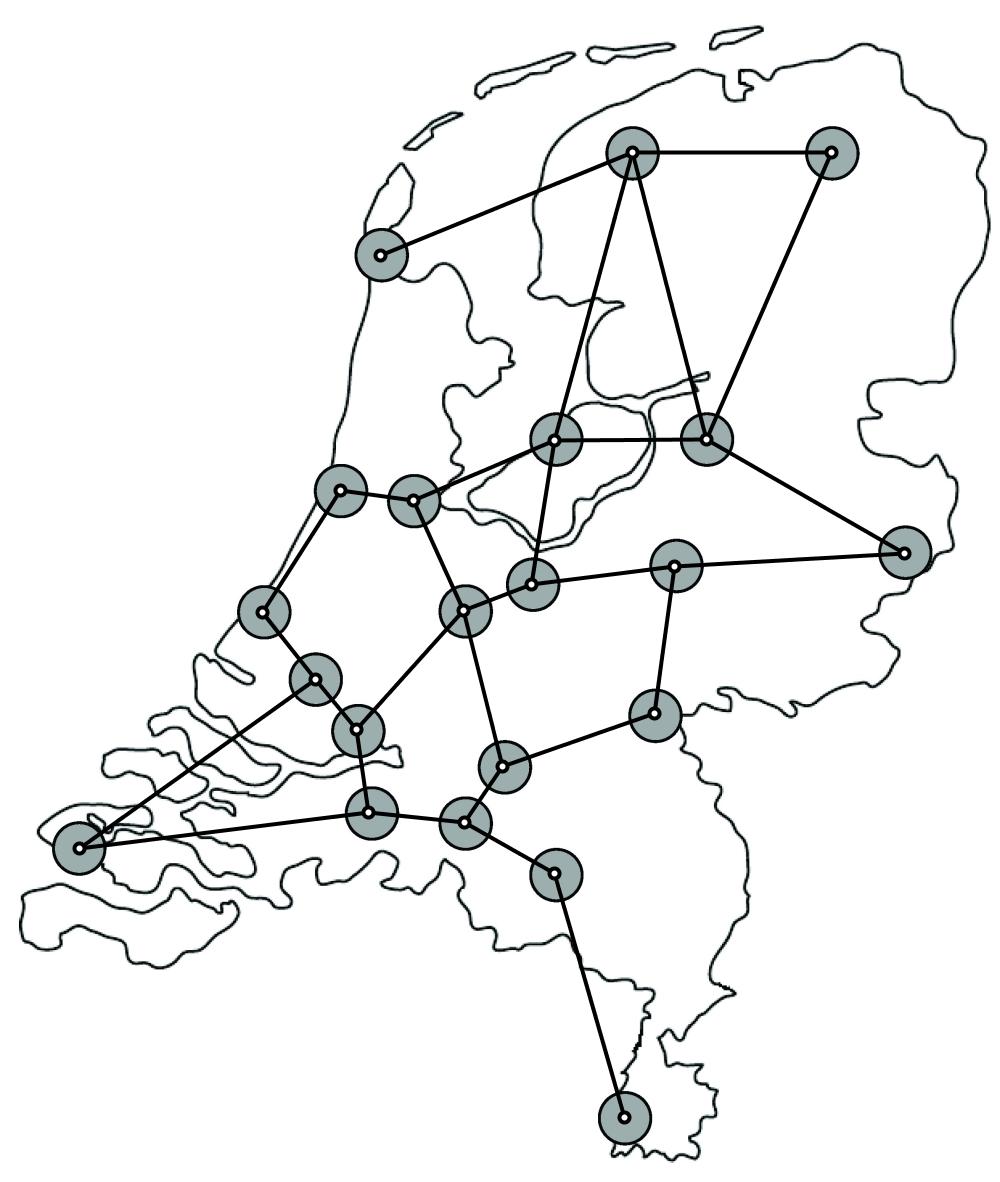}}
\subfigure[$\mathbf P \bigcap \mathbf{ GG}$]{\includegraphics[width=0.32\textwidth]{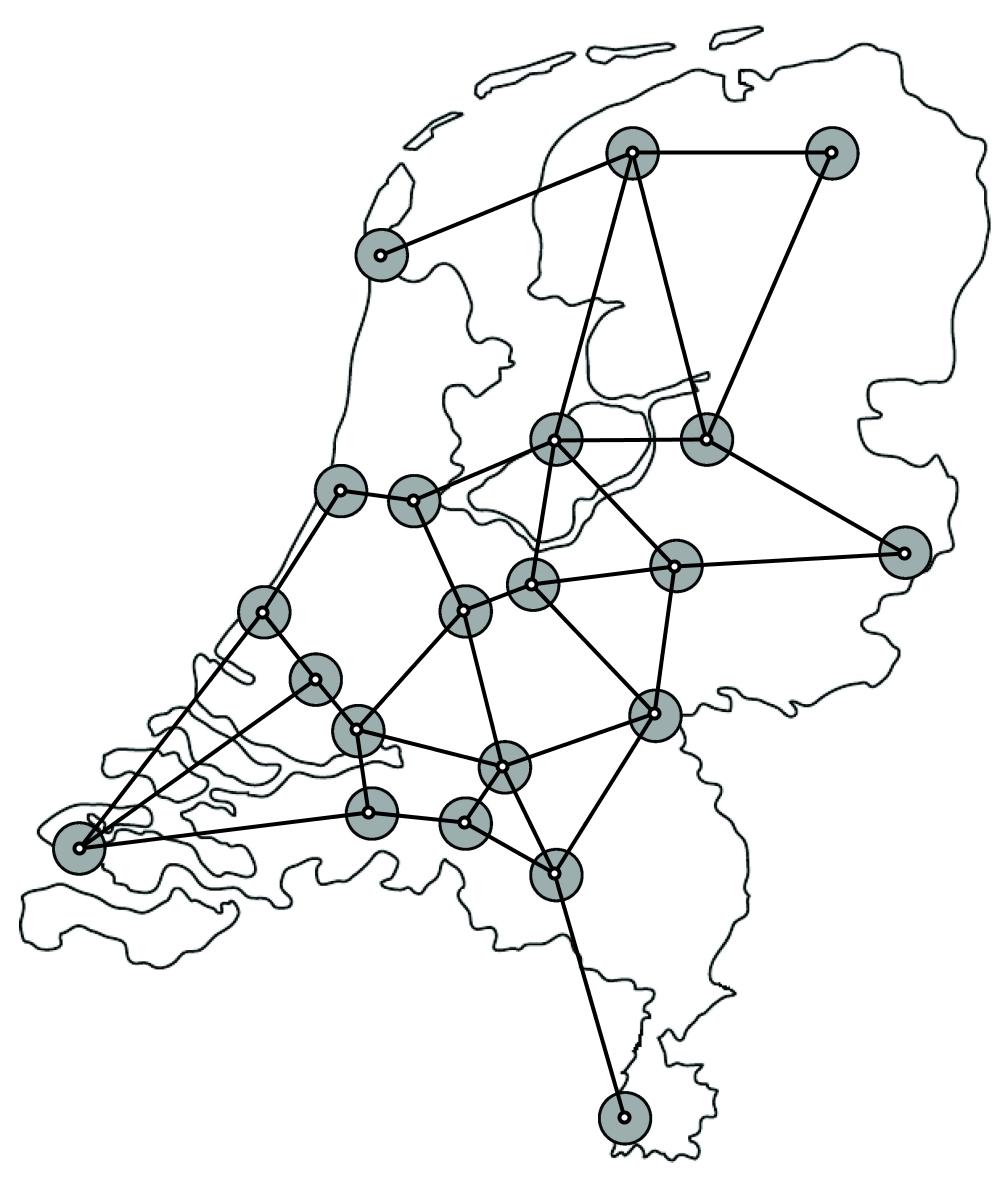}}
\subfigure[$\mathbf P \bigcap \mathbf{ MST}$]{\includegraphics[width=0.32\textwidth]{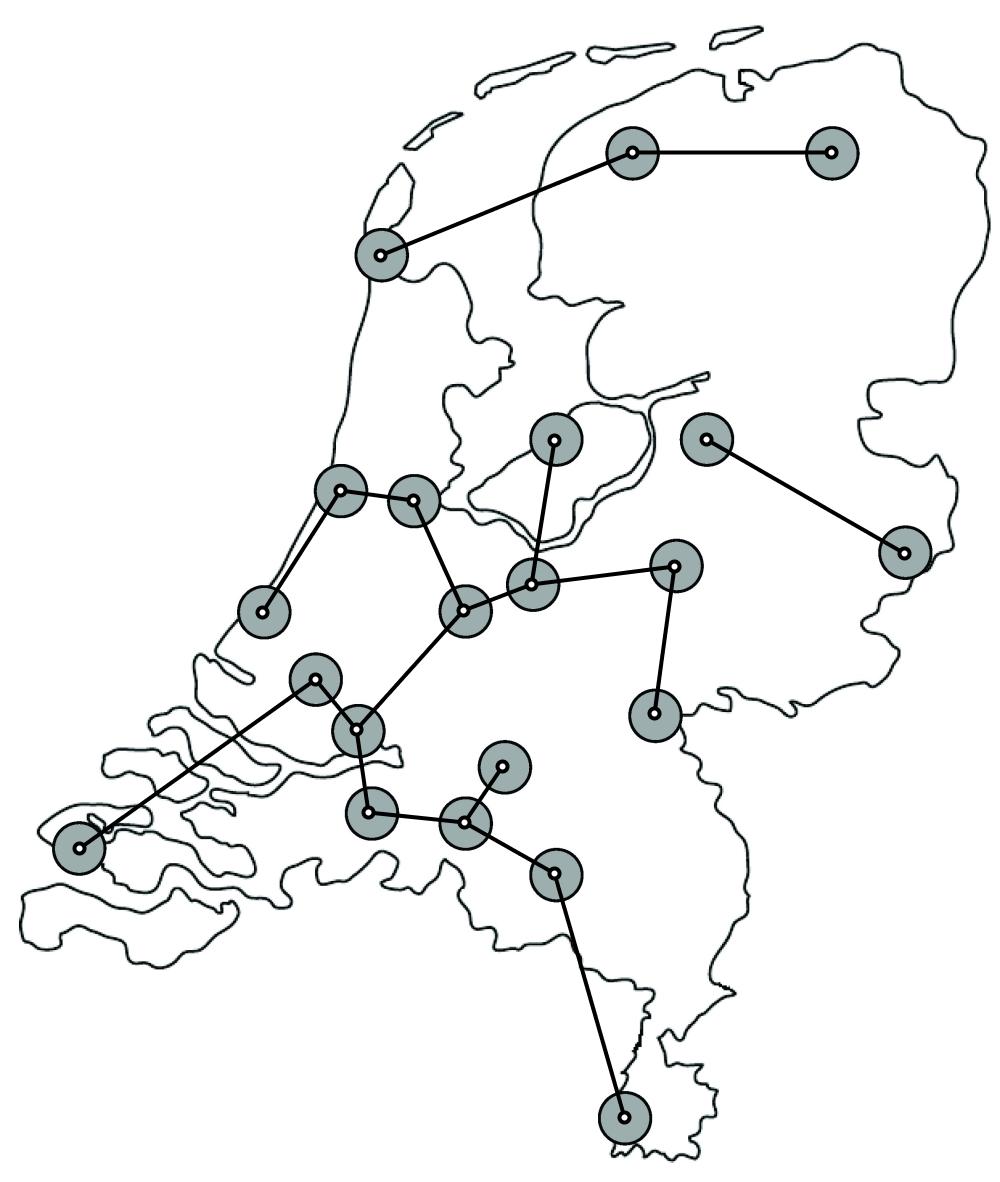}}
\subfigure[$\mathbf H \bigcap \mathbf{ RNG}$ ]{\includegraphics[width=0.32\textwidth]{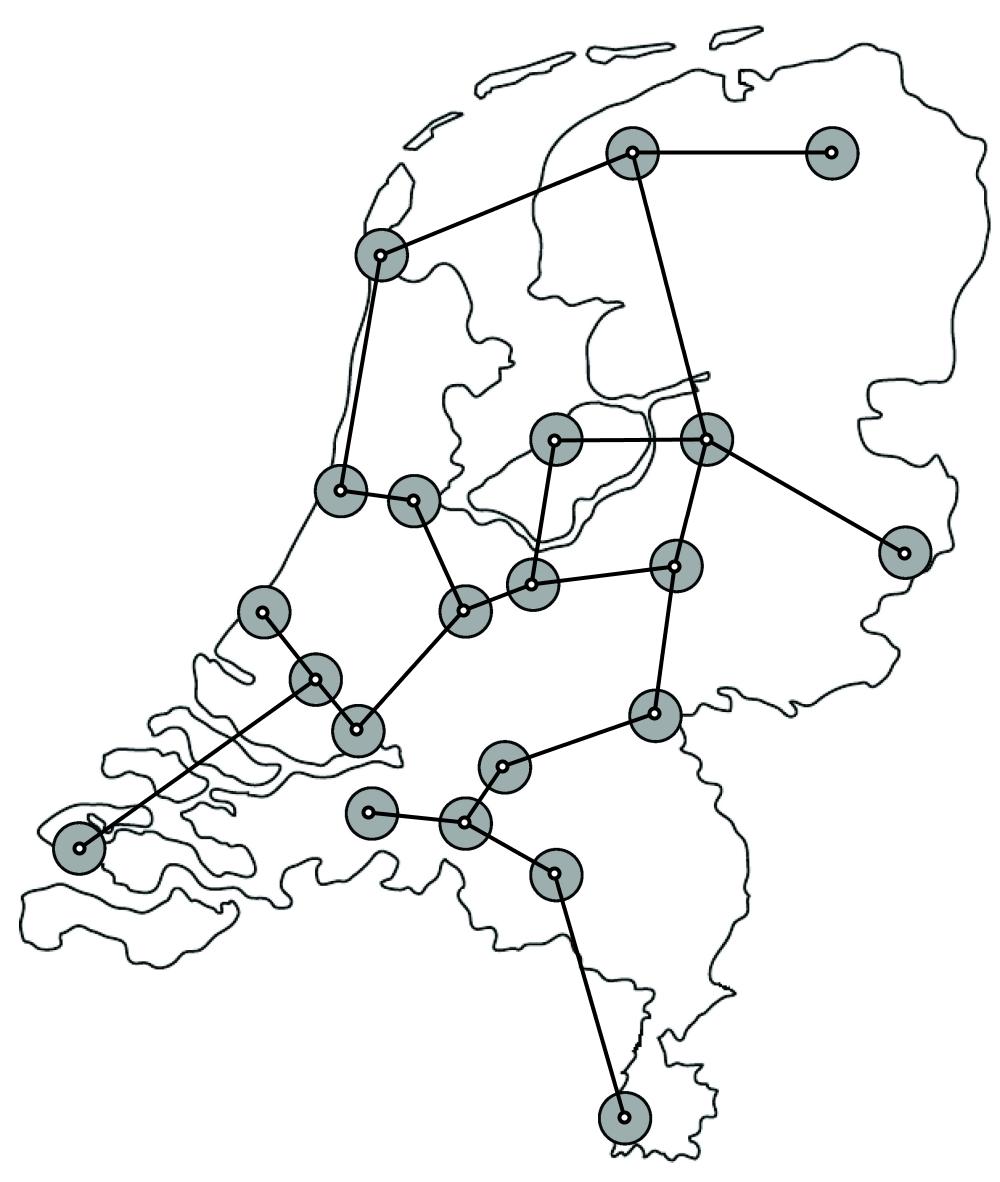}}
\subfigure[$\mathbf H \bigcap \mathbf{ BS}$(1.5)]{\includegraphics[width=0.32\textwidth]{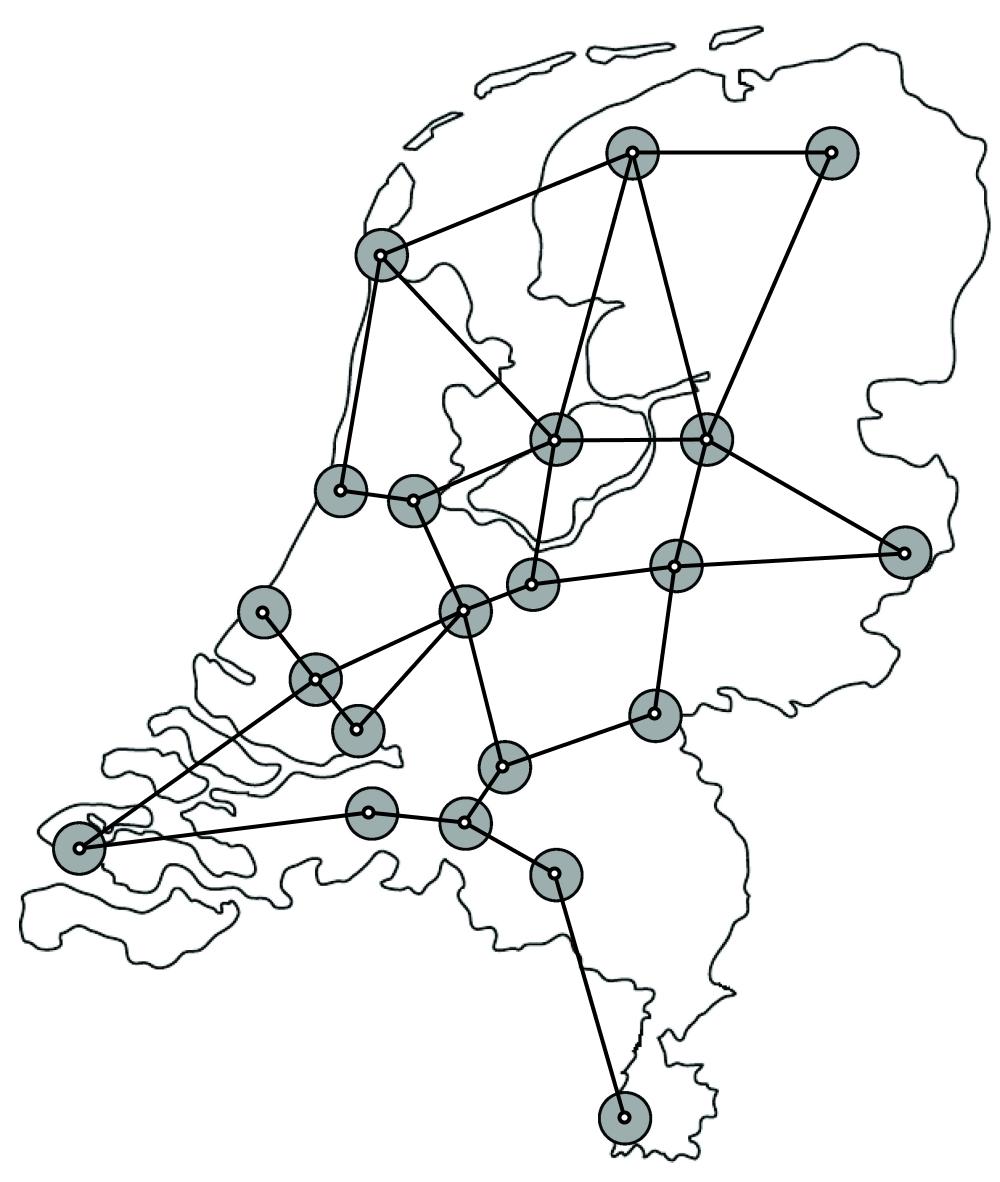}}
\subfigure[$\mathbf H \bigcap \mathbf{ GG}$]{\includegraphics[width=0.32\textwidth]{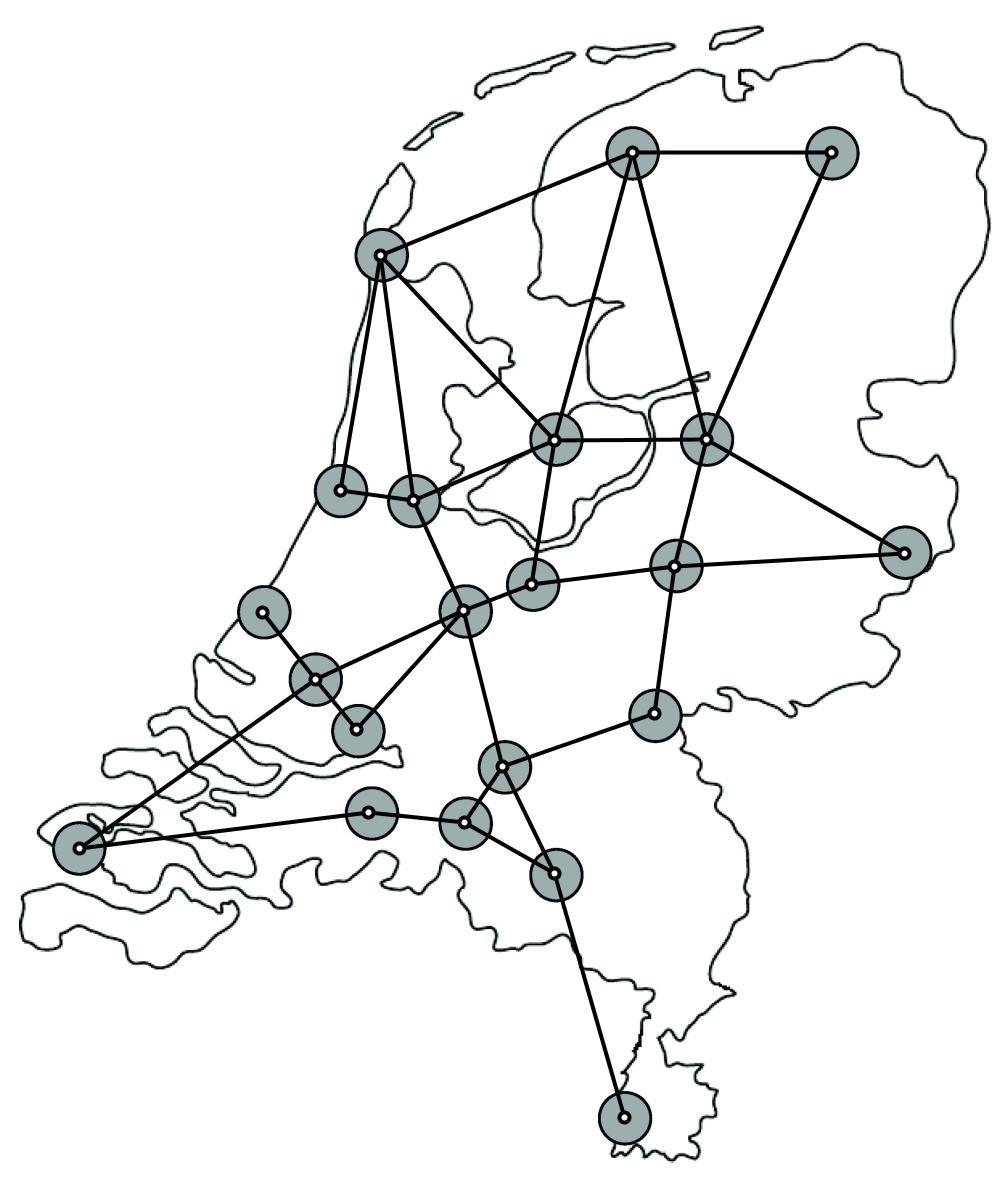}}
\subfigure[$\mathbf H \bigcap \mathbf{ MST}$]{\includegraphics[width=0.32\textwidth]{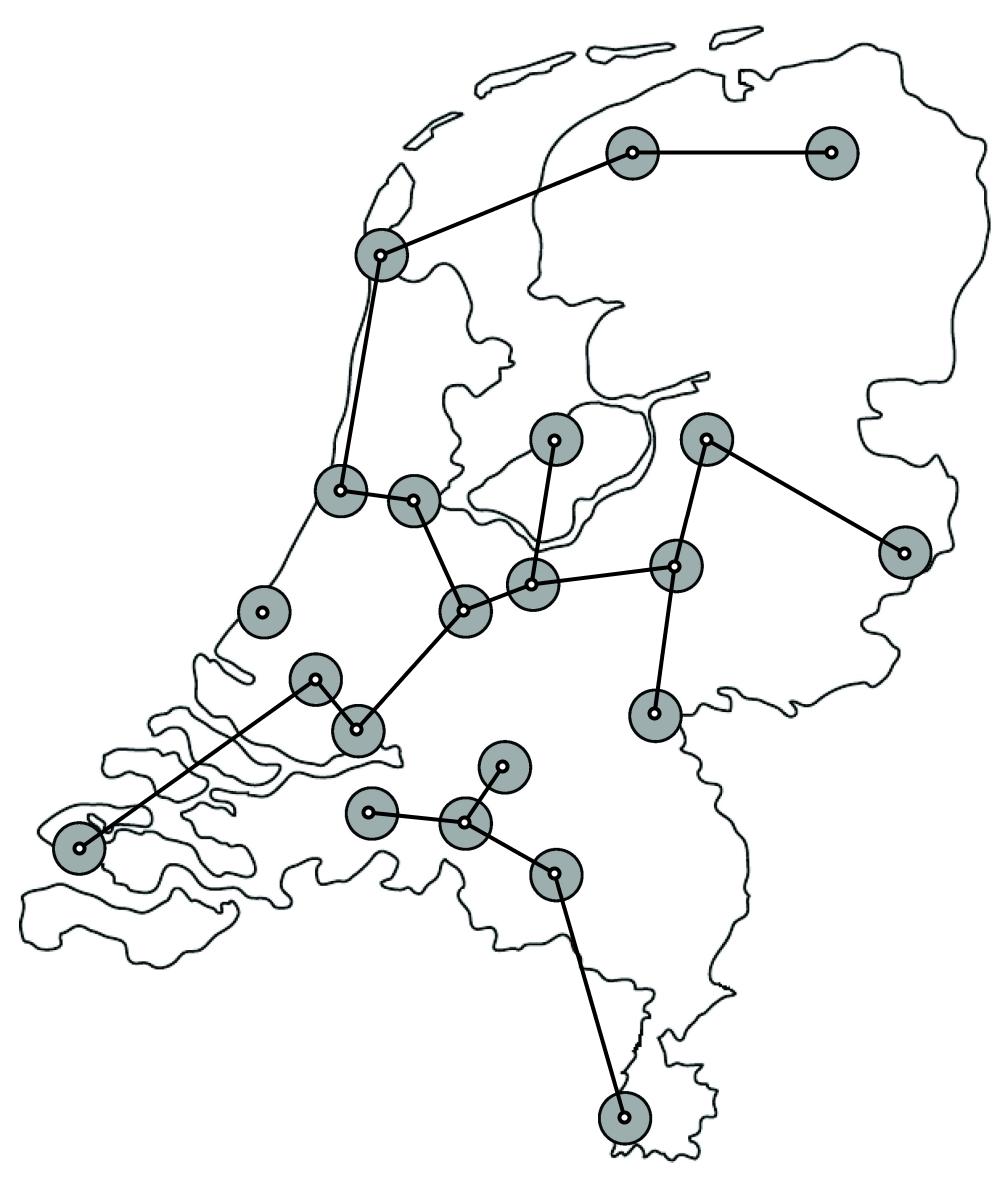}}
\caption{Intersection of Physarum graph $\mathbf P$$(0)$~(a)--(d) and motorway graph $\mathbf H$~(e)--(h) 
with proximity relative neighbourhood graph $\mathbf{ RNG}$, $\beta$-skeleton $\mathbf{ B}$(1.5),
Gabriel graph $\mathbf{ GG}$ and minimum spanning tree $\mathbf{ MST}$.}
\label{intersections}
\end{figure}

We constructed a relative neighbourhood graph~\cite{toussaint_1980}
$\mathbf{RNG}$ (Fig.~\ref{proximity}a), a Gabriel
graph~\cite{gabriel_sokal_1969,matula_sokal_1984} $\mathbf{GG}$
(Fig.~\ref{proximity}c), $\beta$-skeleton ((Fig.~\ref{proximity}b) and a
minimum spanning tree $\mathbf{MST}$ (Fig.~\ref{proximity}d) (we
rooted $\mathbf{MST}$ in Amsterdam) over nodes corresponding to
centres of urban areas. We then calculated intersections of these graphs
with the Physarum graph $\mathbf{P}(0)$ and the motorway graph $\mathbf{H}$,
see Fig.~\ref{intersections}.

The following is a list of edges of the proximity graphs that are not present in Physarum or motorway graphs:
\begin{itemize}
\item $\mathbf P \bigcap \mathbf{MST}$ =$\mathbf{MST}     			- \{$(Den Helder, Haarlem), (Zwolle, Apeldoon)$\}$
\item $\mathbf P \bigcap \mathbf{RNG}$ =$\mathbf{RNG}     			- \{$(Den Helder, Haarlem), (Zwolle, Apeldoon)$\}$
\item $\mathbf P \bigcap \mathbf{BS}(1.5)$ =$\mathbf{BS}(1.5) 	- \{$(Den Helder, Haarlem), (Zwolle, Apeldoon), (Utrecht, Den Haag)$\}$
\item $\mathbf P \bigcap \mathbf{GG}$ =$\mathbf{GG} 						- \{$(Den Helder, Haarlem), (Zwolle, Apeldoon), (Utrecht, Den Haag), (Den Helder, Amsterdam), (Den Helder, Lelystad)$\}$

\item $\mathbf H \bigcap \mathbf{MST}$ =$\mathbf{MST}     			- \{$(Haarlem, Enschede), (Rotterdam, Hertogenbosch)$\}$
\item $\mathbf H \bigcap \mathbf{RNG}$ =$\mathbf{RNG}     			- \{$(Haarlem, Enschede), (Rotterdam, Hertogenbosch)$\}$
\item $\mathbf H \bigcap \mathbf{BS}(1.5)$ =$\mathbf{BS}(1.5) 	- \{$(Haarlem, Enschede), (Rotterdam, Hertogenbosch)$\}$
\item $\mathbf H \bigcap \mathbf{GG}$ =$\mathbf{GG} 						- \{$(Haarlem, Enschede), (Rotterdam, Hertogenbosch), (Enschede, Tilburg), (Rotterdam, Nijmegen), (Amersfoort, Dordrecht), (Dordrecht, Tilburg)$\}$.
\end{itemize}

The motorway graph closely matches the spanning tree, relative
neighbourhood graph and $\beta$-skeleton.  Only two edges {(Haarlem,
  Enschede), (Rotterdam, Hertogenbosch)}, presented in $\mathbf{RNG}$,
$\mathbf{MST}$, $\mathbf{BS}(1.5)$ do not exist in $\mathbf H$. The
fact that the relative neighbourhood graph is \emph{almost} a
sub-graph of the motorway graph indicates intrinsically logical
organization of the transport networks in the Netherlands. This is
because a relative neighbourhood graph is commonly considered to be
optimal in terms of total edge length and travel distance, and is
known to be a good approximation of road networks~\cite{watanabe_2005,watanabe_2008}.

Said that the transport networks in the Netherlands
are redundant --- from slime mould's point of view --- because there is a substantial number of edges of
$\mathbf H$ not presented in $\mathbf{RNG}$.

The same can be said about the Physarum graph, because only edges
(Den Helder, Haarlem) and (Zwolle, Apeldoon) of $\mathbf{RNG}$ are not
represented by protoplasmic tubes. Under-representations of
$\beta$-skeleton and $\mathbf{GG}$ in between $\mathbf{P}(0)$ are much
more substantial: three and five edges, respectively.

\section{Flooding}

\label{flooding}

\begin{figure}[!tbp]
\centering
\subfigure[]{\includegraphics[width=0.49\textwidth]{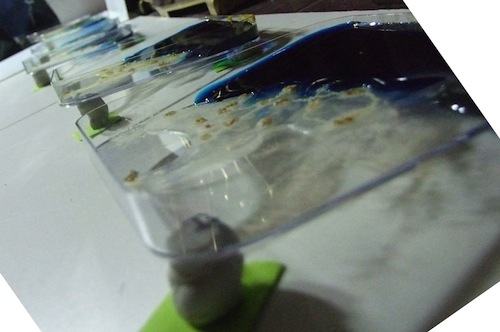}}
\subfigure[]{\includegraphics[width=0.49\textwidth]{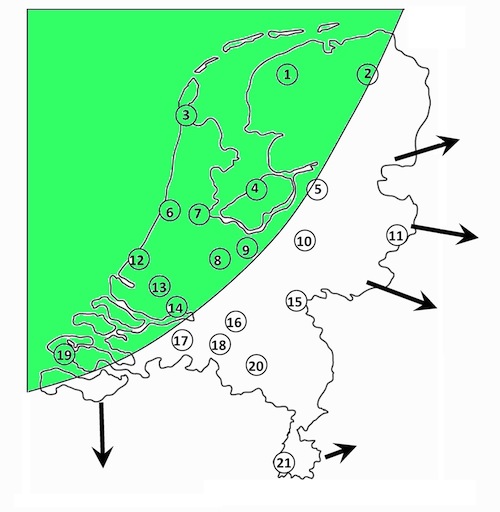}}
\caption{Flooding setup: (a)~array of Petri dishes during imitated flooding, 
(b)~Flooding scheme. Flooded part is filled is shaded (green), direction of outward 
migration are shown by arrows, each large arrow represents 20\%, while small arrow 10\%. }
\label{floodingscheme}
\end{figure}

In this section we decribe the experiments which investigate the
capability of the plasmodium to adapt the its network during induced
flooding of the petri-dish. In essence we investigate how the
plasmodium would calculate and adapt the transport network, if the
Netherlands were to suffer similar flooding of its cities and roads.
Experiments on flooding were conducted in 12$\times$12~cm Petri
dishes. A Petri dish was raised by 1-2~cm in its south-east corner and
partly filled with liquid (distilled water, either pure or coloured
with one drop of food coloring) (Fig.~\ref{floodingscheme}a). In most
cases the flooded area included Middelburg on the south-west and
Groningen on the north-east. In the central part of the country the
flooding often reached Zwolle. The exact flooded area varied between
experiemtns due to slight variations in the thickness of agar gel
substrate and minor differences in inclinations of Petri dishes,
the flooding scheme shown in (Fig.~\ref{floodingscheme}) is rather
indicative.

\begin{figure}[!tbp]
\centering
\subfigure[$t=0$~h]{\includegraphics[width=0.49\textwidth]{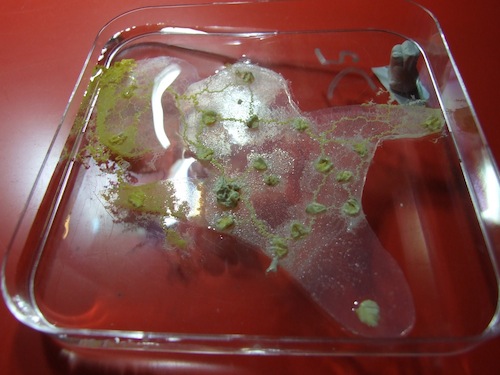}}
\subfigure[$t=30$~h]{\includegraphics[width=0.49\textwidth]{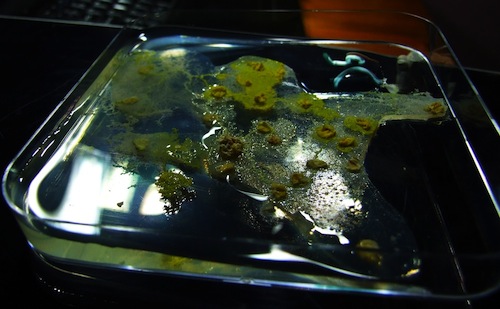}}
\subfigure[$t=46$~h]{\includegraphics[width=0.49\textwidth]{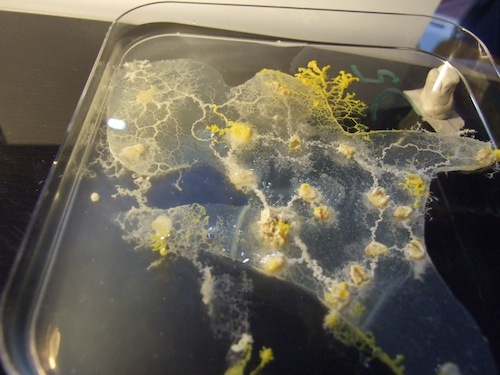}}
\caption{Illustration of plasmodium behaviour during experiment on flooding. Photographs 
are made at different angles. }
\label{flooding1}
\end{figure}

Initially plasmodium reacts to flooding with increased activity.
During the first few hours of flooding the plasmodium typically
increases its branching at the boundary of the flooded area
(Fig.~\ref{flooding1}a). Often there are indications of indiscriminate
increase of foraging, panic foraging. For example, in
Fig.~\ref{flooding1}b we can see active sprawling of plasmodium in the
areas around Apeldoon, Dordrecht and Enschede: no protoplasmic tubes
are formed but rather uniform sheets of plasmodium propagate in these
areas. Eventually the activity ceases and flooded transport links
becomes abandoned (Fig.~\ref{flooding1}c).

\begin{figure}[!tbp]
\centering
\subfigure[]{\includegraphics[width=0.49\textwidth]{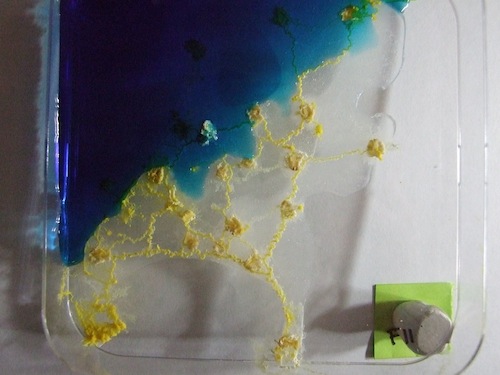}}
\subfigure[]{\includegraphics[width=0.49\textwidth]{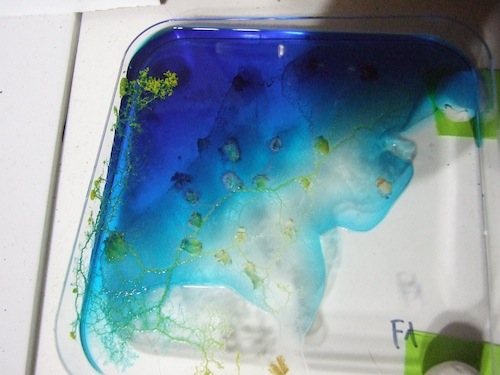}}
\caption{Compensation of transport networks (a) and increase in borderline transport (b). }
\label{increasedactivityalongroads}
\end{figure}

In some cases no `panic' branching occurs, but non-flooded
protoplasmic tubes become thicker due to increased transport of
nutrients and relocation of masses of protoplasm from areas affected
by flooding (Fig.~\ref{increasedactivityalongroads}a). Often only
protoplasmic tubes located along flood line are hypertrophied, for
example there is an increased thickness of tubes along the route
Hertogenbosch - Apeldoon - Zwolle - Groningen in
Fig.~\ref{increasedactivityalongroads}b.

\begin{figure}[!tbp]
\centering
\subfigure[]{\includegraphics[width=0.49\textwidth]{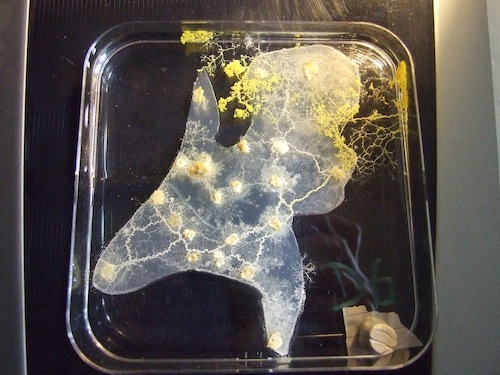}}
\subfigure[]{\includegraphics[width=0.49\textwidth]{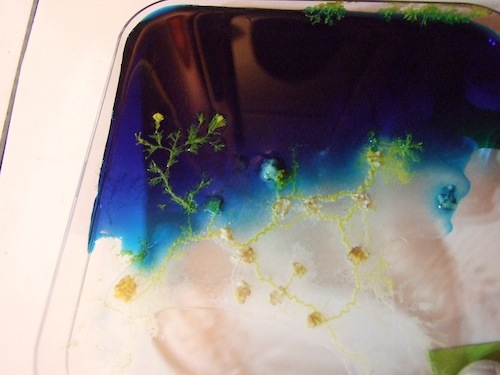}}
\caption{Examples of evacuation.}
\label{floating}
\end{figure}

With time, water is absorbed by the gel and is sucked under the gel plate
due to capillary forces, which in turn causes the overall humidity to
increase.  This is associated with a reduced concentration of nutrients,
and an increased concentration of metabolites ejected in the agar plate,
force the plasmodium to abandon the agar plate and migrate
beyond. Often the plasmodium attempts to complete evacuation by
crawling onto the water surface (for example see Fig.~\ref{floating}).

Outward migration, calculated in nine experiments, are
shown in Fig.~\ref{floodingscheme}b. We conclude that if a flooding of
the Netherlands were to happen the major impact of migration (60\%) will be
felt by Western Germany, with just a minor impact on Northen France (20\%)
and a very slight impact on Belgium (10\%). If the Netherlands were flooded the 
Western Germany accepts the largest impact of mass-migration.

\section{Discussion}
\label{discussion}

In laboratory experiments with plasmodium of \emph{Physarum
  polycephalum} we discovered that the Physarum protoplasmic network forms
a sub-network of man-made motorway networks, i.e. every transport link
represented by Physarum can also be found as a segment of the motorway
network. However, the converse does not hold. The motorway network is
not a sub-graph of the Physarum network; there are edges of motorway
graph not represented by edges Physarum graph. Transport links
Amsterdam to Der Helden, Zwolle to Apeldoon, and Rotterdam to
Dordrecht are never presented in Physarum graphs. This may be
interpreted as if the man-made motorway network in the Netherlands is
redundant from Physarum point of view.

Also, in many experiments Den
Helder city remains disconnected from other cities. This is either a
mishap of the Physarum approach, e.g. plasmodium prefers not to enter
narrow peninsula of North Holland, or an indication of a somewhat inefficient
location of Den Helder.

The most robust component of the Physarum graph, the component which is
present in the majority of experiments is a tree with three linear
branches. First branch is Tilburg - Breda - Dordrecht - Rotterdam -
Den Haag, second is Tilburg - Hertogenbosch - Nijmegen - Apeldoon, and
third is Tilburg - Eindhoven - Maastricht. A possible explanation
would be that relative positions and distances between the cities in
this trees is optimal for plasmodium physiological functioning. Cities
are not close enough to be contaminated by products of plasmodium
activity but close enough not to put significant strain on the pumping
of nutrients between distant parts of plasmodium's body.

If we prune the Physarum graph by removing edges which occur in less
thana quarter of experiments and look at the intersection (i.e. the set of
edges present in both graphs) of this graph with a graph of motorways
we find that intersection consist of three disconnected
components. The first component is a chain Enschede - Den Haag -
Rotterdam. The second component is a cycle Nijmegen - Breda -
Middelburg with branches Breda - Hertogenbosch - Tilburg, Middelburg -
Eindhoven and Nijmegen - Dordrecht - Apeldoon - Amersfoort. FInally,
the third component is a cycle Leeuwarden - Groningen - Zwolle with
branches Zwolle - Enschede and Zwolle - Lelystad - Amsterdam -
Haarlem, Utrecht. With respect to the proximity graph the key finding
is that relative neighbourhood graphs (which is commonly recognised as
a best approximation of urban streets and transport networks) are
almost (apart of two edges) subgraphs of the Physarum graph and motorway
graph.

By physically imitating flooding of some parts of the Netherlands we
predicted that if a real flooding were to occur, the following events will
take place: substantial increase in traffic on the parts of motorway
networks close to the boundary between flooded and non-flooded areas,
propagation of the traffic congestion to all non-flooded parts of the
country, complete paralysis and abandonment of transport network,
migration of population from the Netherlands to Germany and France,
and Belgium.

\begin{figure}[!tbp]
\centering
\subfigure[$t=1$]{\includegraphics[width=0.32\textwidth]{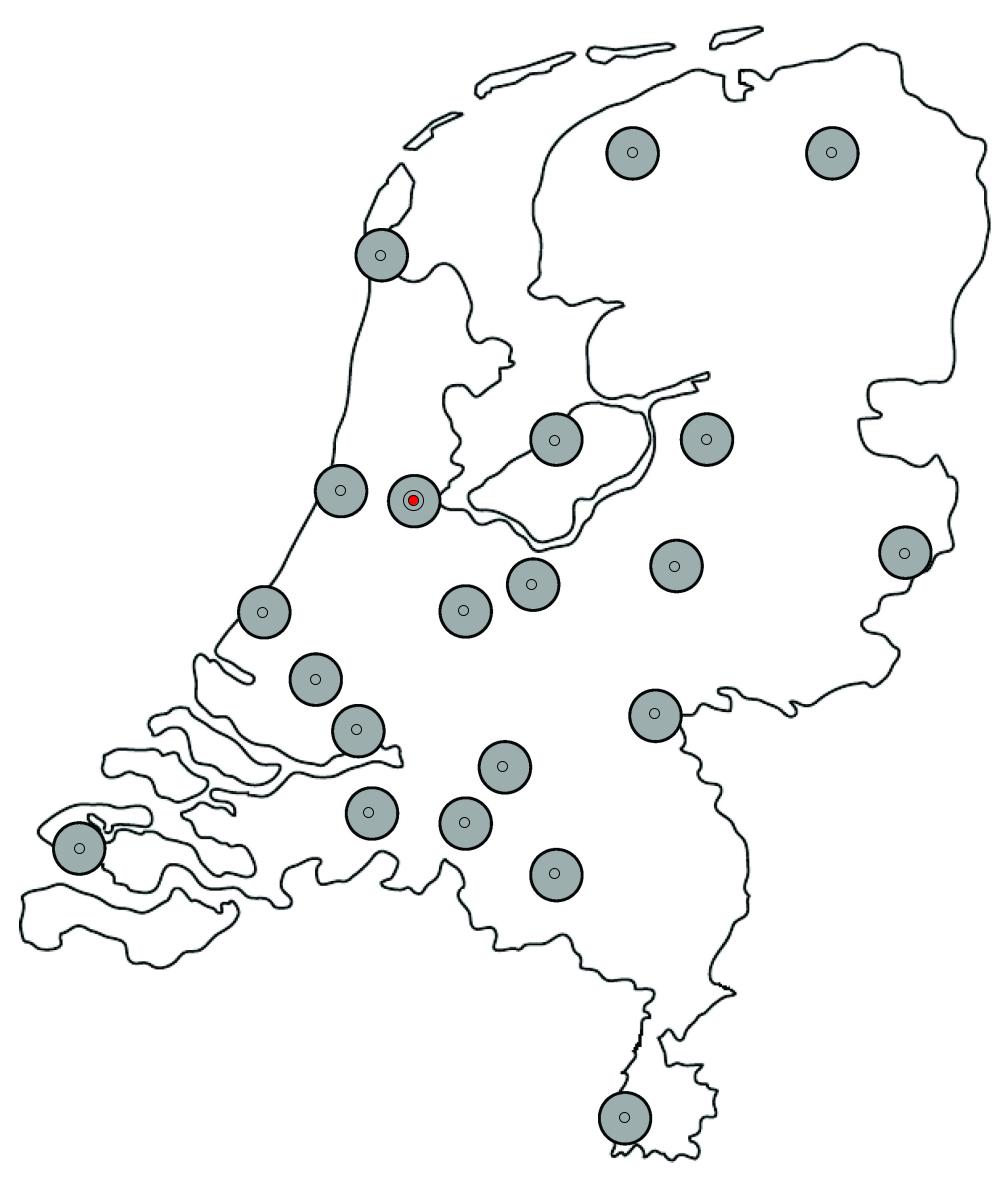}}
\subfigure[$t=2$]{\includegraphics[width=0.32\textwidth]{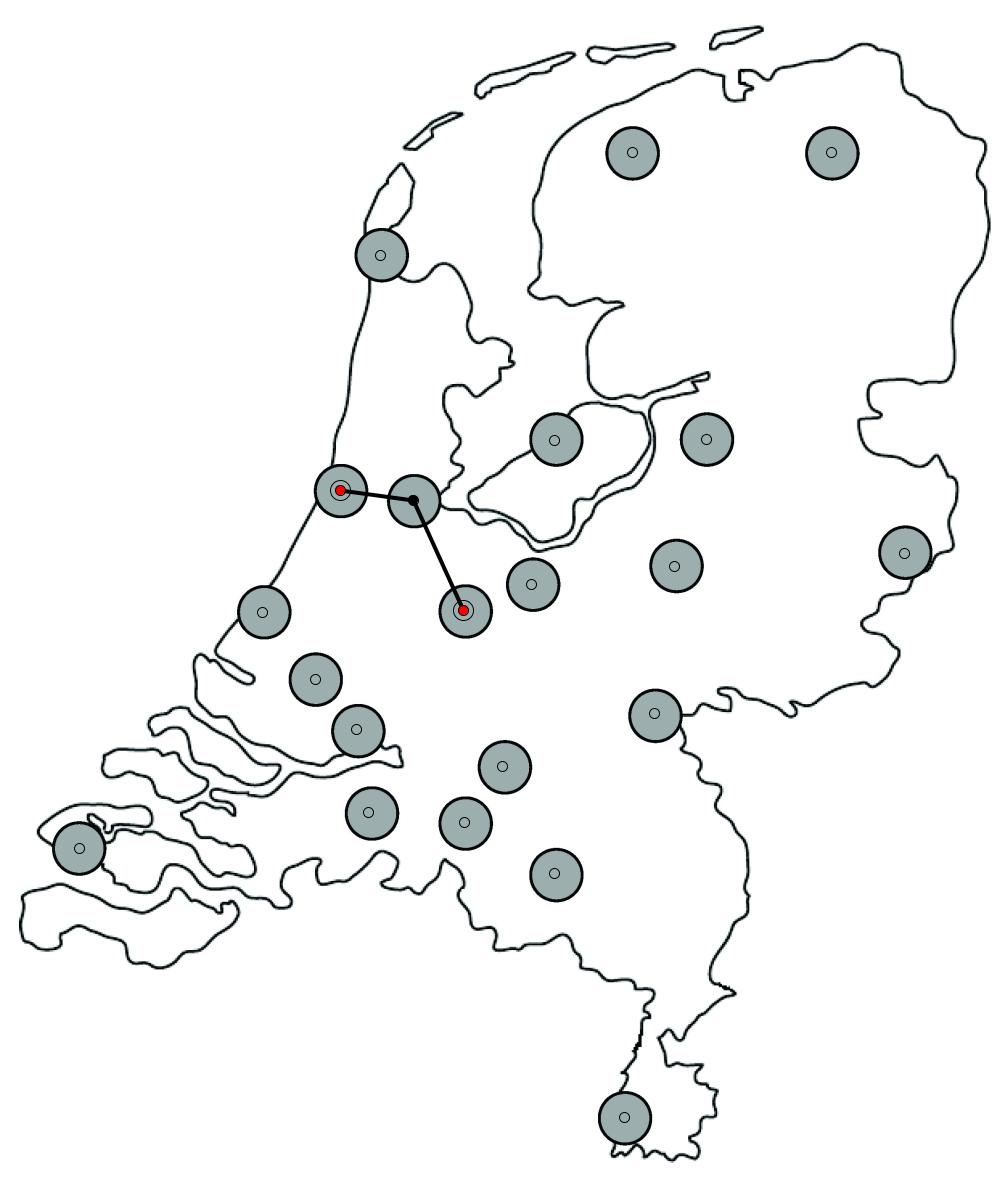}}
\subfigure[$t=3$]{\includegraphics[width=0.32\textwidth]{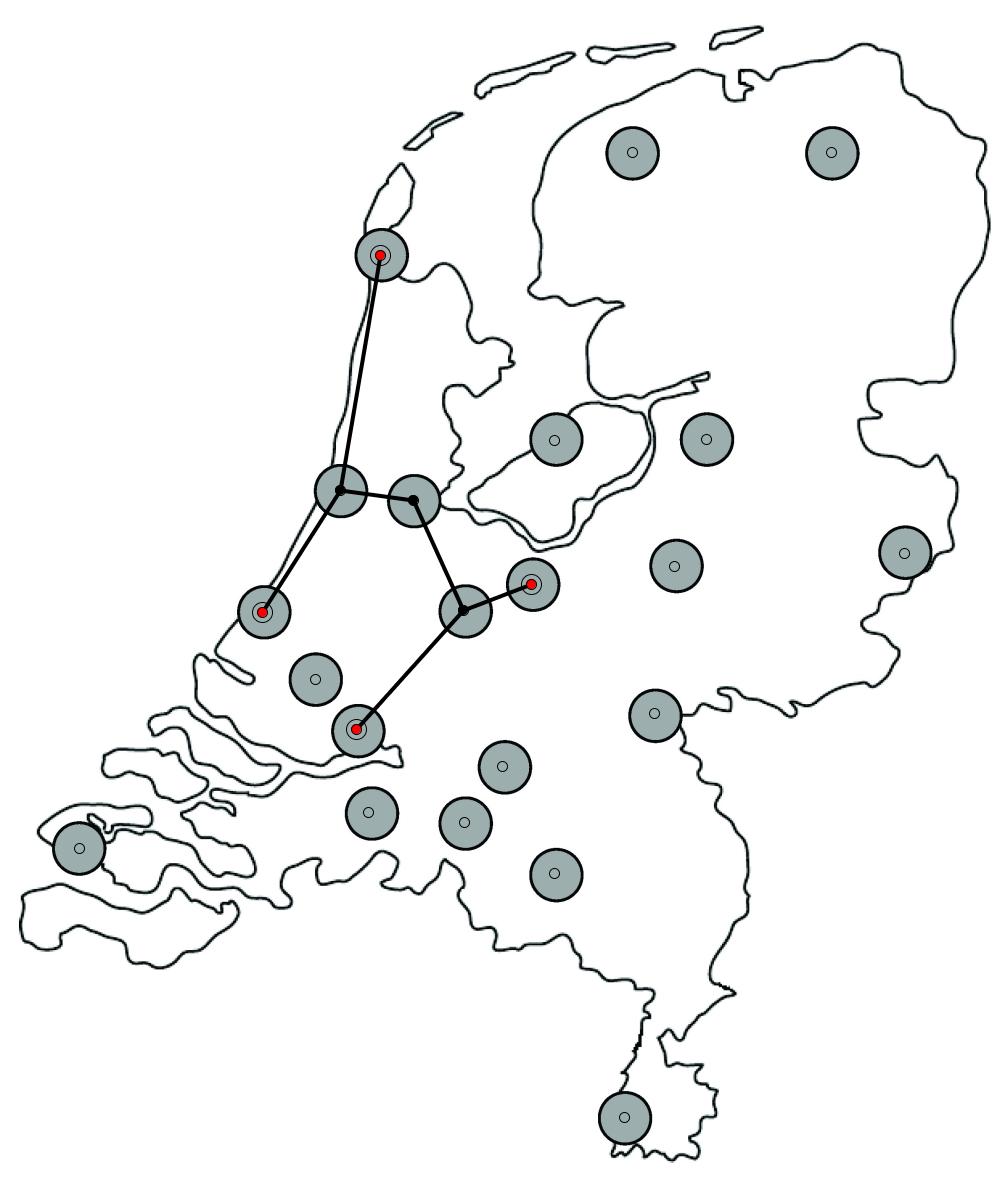}}
\subfigure[$t=4$]{\includegraphics[width=0.32\textwidth]{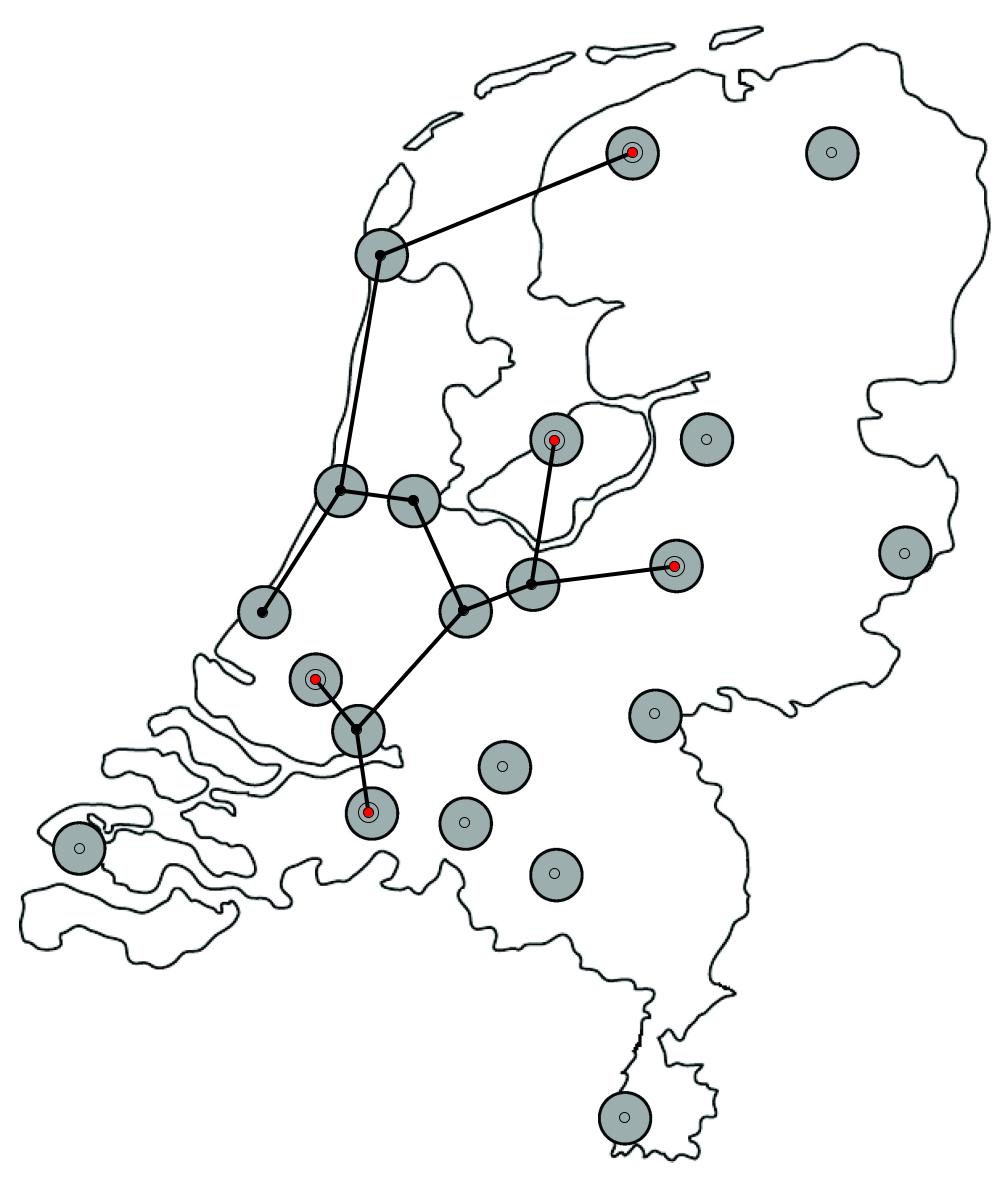}}
\subfigure[$t=5$]{\includegraphics[width=0.32\textwidth]{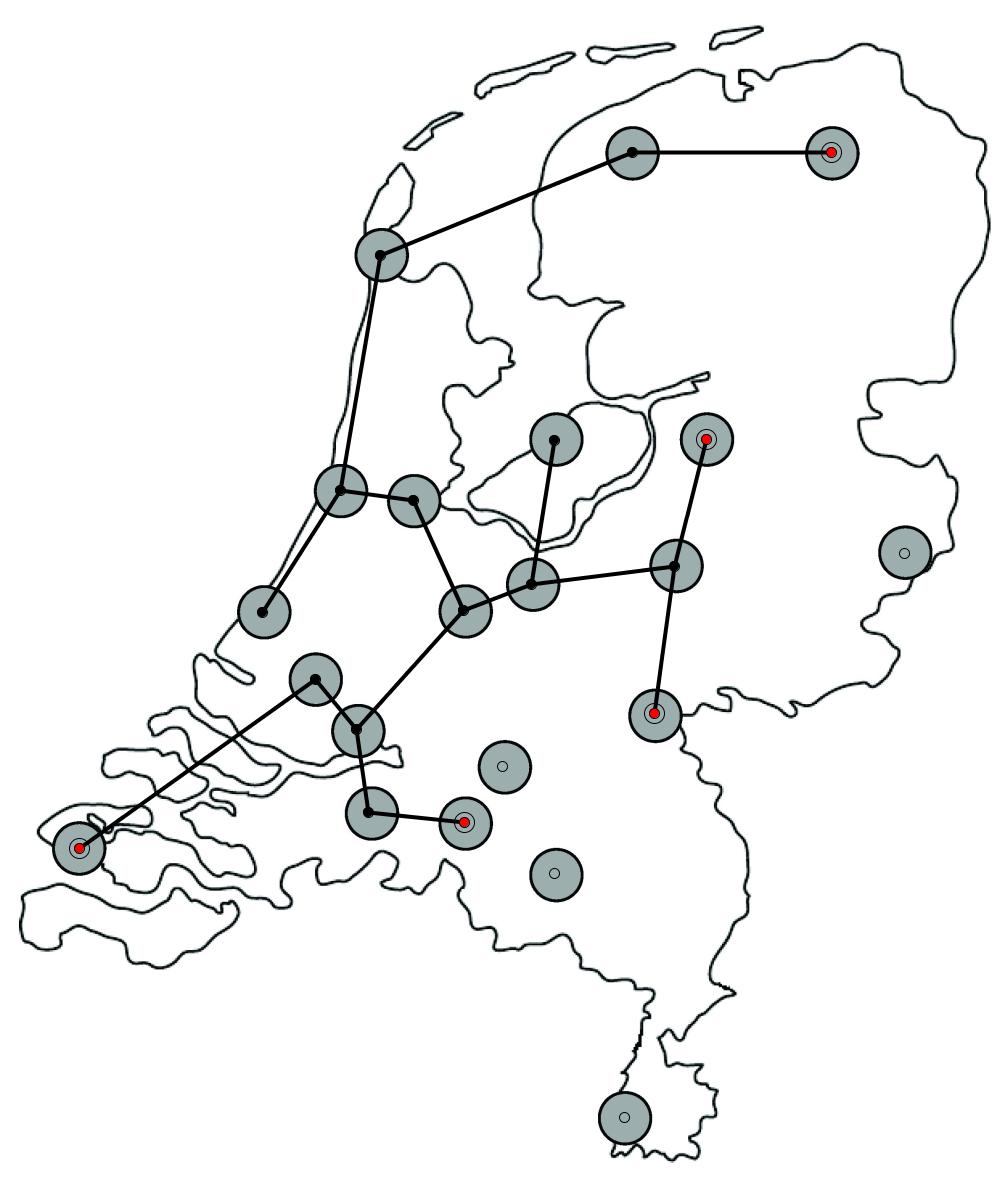}}
\subfigure[$t=7$]{\includegraphics[width=0.32\textwidth]{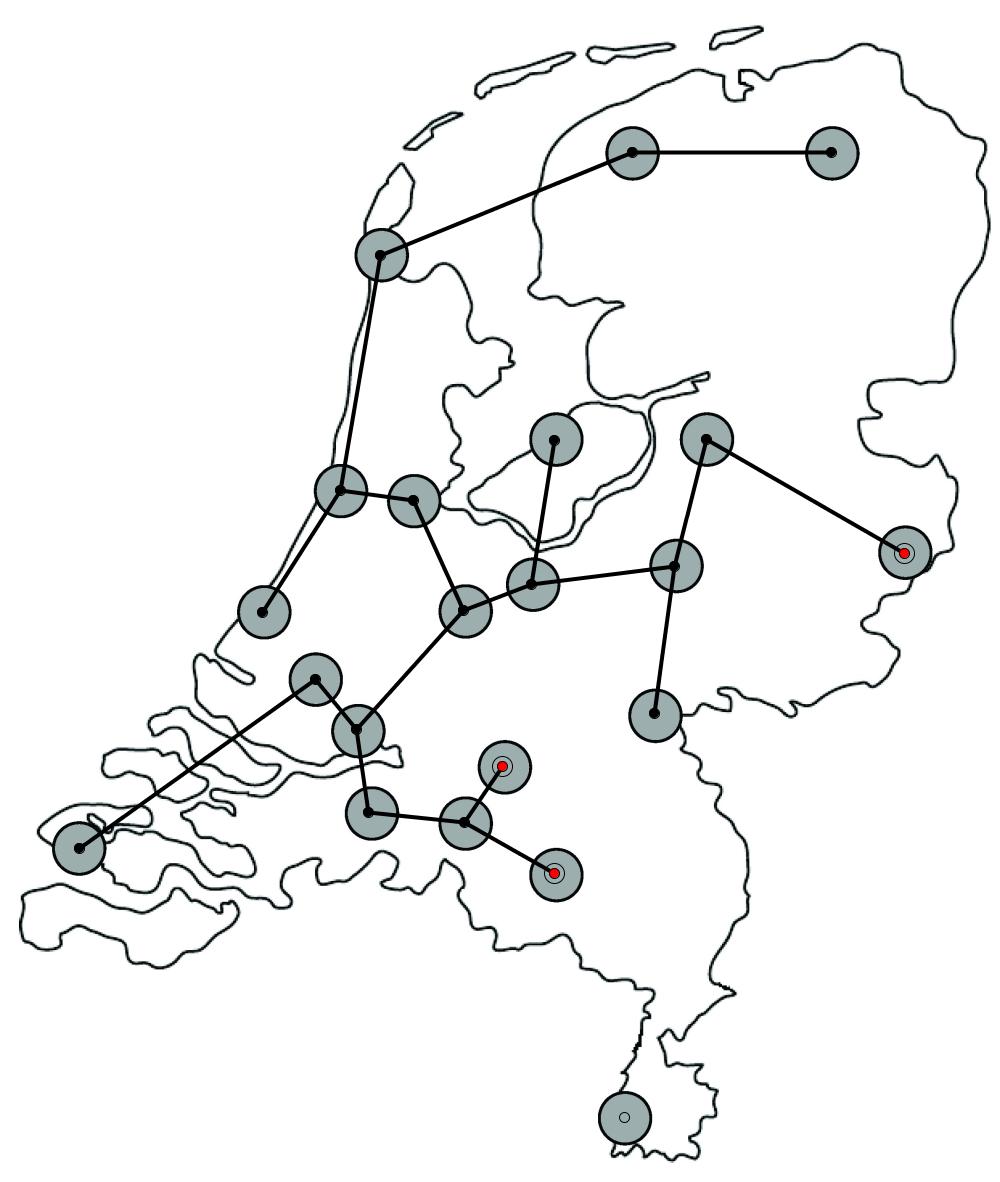}}
\subfigure[$t=8$]{\includegraphics[width=0.32\textwidth]{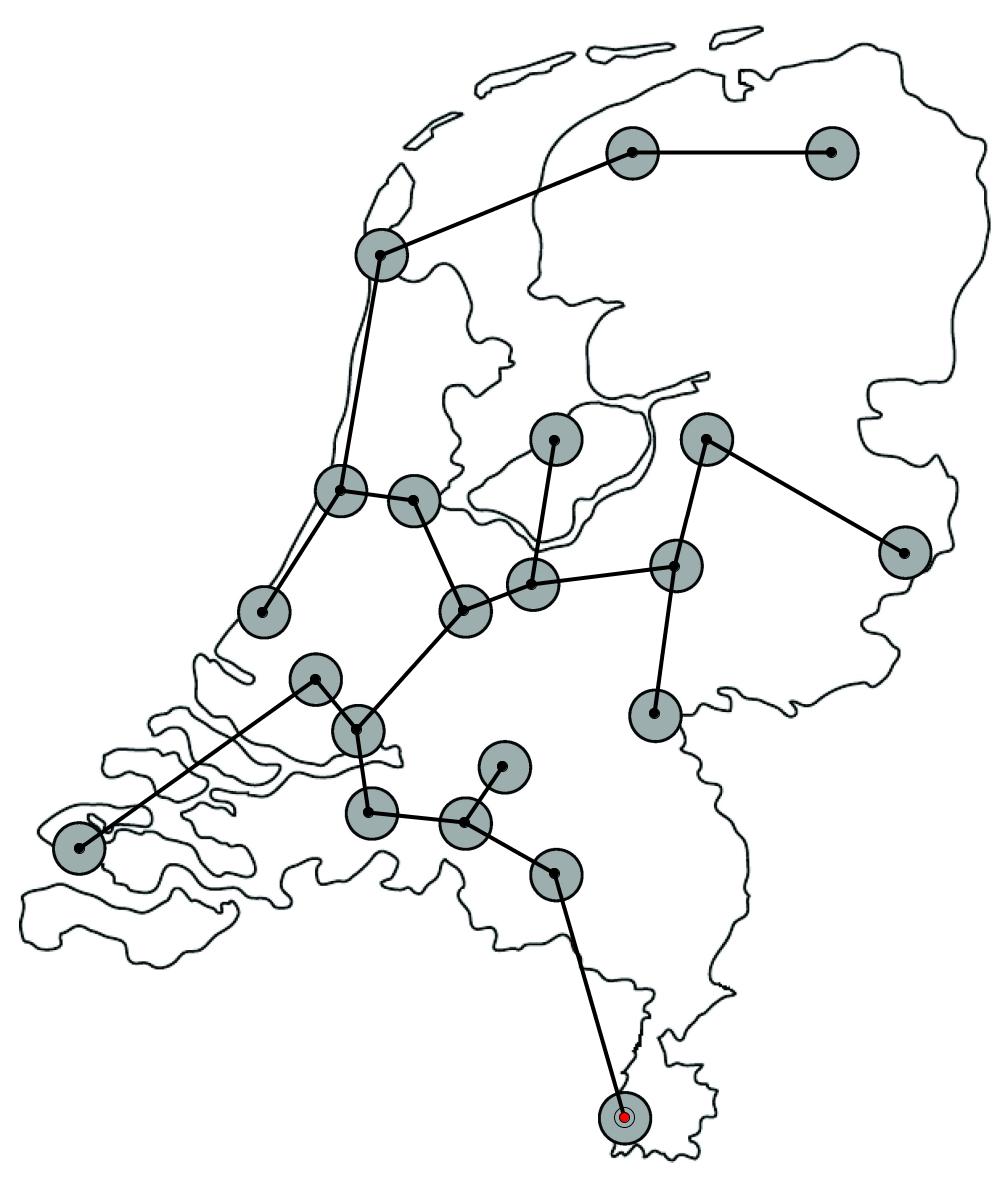}}
\subfigure[$t=9$]{\includegraphics[width=0.32\textwidth]{figs/TreeGrows/0011}}
\caption{Snapshots of a growing spanning tree of $\mathbf U$ rooted in Amsterdam. Vertices active at time step $t'$ are shown by 
black disc at snapshot $t=t'$.}
\label{treegrowth}
\end{figure}

The plasmodium is incredibly similar, esp. in wave-like behaviour, to
sub-excitable non-linear
media~\cite{adamatzky_naturewissenschaften_2007} and is mainly guided
by gradients of chemo-attractants~\cite{adamatzky_bz_trees}. Based on
these two facts we could assume the plasmodium must spread from
Amsterdam omni-directionally and then start forming branches between
cities, as e.g. illustrated in Fig.~\ref{treegrowth}. Such phenomena do
not occur, rather the plasmodium behaves more like a `single-headed'
creature, it chooses one direction of movement, explores it, then
chooses another direction and explores it again, see
Fig.~\ref{arrowsf12}. Such behaviour of the plasmodium may explain the
differences between ideal planar proximity graphs and protoplasmic
networks constructed by the plasmodium.

In future work we plan to undertake more experiments on Physarum-based
imitation of road formation in other European countries, and may be
even make experiments at a large scale, where plasmodium grows over
the whole Europe.  We also plan to evaluate the city state of
Singapore, this present a fairly unique road network that has been
carefully planned and constructed within the last 50 years. One
further objective would be to try and incorporate three dimensional
landscape elevation into our laboratory experiments.
'

\section*{Acknowledgements}

Peter M.A. Sloot would like to acknowledge the support of this work by the
Russian Federation Leading Scientist Grant, contract: 11.G34.31.0019
and the European Union Dynanets project: 233847.

\end{document}